\documentclass[letterpaper]{article} 
\usepackage{aaai25}  
\usepackage{times}  
\usepackage{helvet}  
\usepackage{courier}  
\usepackage[hyphens]{url}  
\usepackage{graphicx} 
\usepackage{natbib}  
\usepackage{caption} 
\usepackage{algorithm}
\usepackage[noend]{algorithmic}

\usepackage{newfloat}
\usepackage{listings}
\DeclareCaptionStyle{ruled}{labelfont=normalfont,labelsep=colon,strut=off} 
\floatstyle{ruled}
\newfloat{listing}{tb}{lst}{}
\floatname{listing}{Listing}
\title{Improving Community-Participated Patrol for Anti-Poaching}
\author {
    Yufei Wu\textsuperscript{\rm 1},
    Yixuan Even Xu\textsuperscript{\rm 2},
    Xuming Zhang\textsuperscript{\rm 3},
    Duo Liu\textsuperscript{\rm 3},
    Shibing Zhu\textsuperscript{\rm 4},
    Fei Fang\textsuperscript{\rm 2}
}
\affiliations {
    \textsuperscript{\rm 1}Shanghai Jiao Tong University\\
    \textsuperscript{\rm 2}Carnegie Mellon University\\
    \textsuperscript{\rm 3}World Wide Fund For Nature China\\
    \textsuperscript{\rm 4}Heilongjiang Academy of Sciences\\
    yvonne2021@sjtu.edu.cn, yixuanx@cs.cmu.edu, uriel900401@gmail.com, \\
    dliu@wwfchina.org, zhushibing@haszrs.org.cn, feif@cs.cmu.edu
}

\usepackage{enumerate}

\usepackage{subcaption}

\usepackage{amsthm}
\usepackage{amsmath}
\usepackage{amssymb}
\usepackage{mathtools}

\usepackage[capitalise]{cleveref}

\usepackage{threeparttable}
\usepackage{hhline}
\usepackage{multirow}

\usepackage{tikz}
\usetikzlibrary{calc}

\usepackage{xcolor}

\usepackage{thmtools}
\usepackage{thm-restate}

\newtheorem{definition}{Definition}[section]

\newtheorem{lemma}{Lemma}[section]

\newenvironment{proofof}[1]{{\bf \noindent Proof of #1:  }}{\hfill\rule{2mm}{2mm}}

\renewcommand{\b}{\mathbf}
\newcommand{\mr}{\mathrm}

\newcommand{\I}{\mathcal I}
\newcommand{\rp}{r^{\mathrm{p}}}
\newcommand{\rv}{r^{\mathrm{v}}}
\newcommand{\ep}{e^{\mathrm{p}}}
\newcommand{\ev}{e^{\mathrm{v}}}
\newcommand{\ud}{U^{\mathrm{d}}}
\newcommand{\udp}{U^{\mathrm{d}\prime}}
\newcommand{\ua}{U^{\mathrm{a}}}
\newcommand{\uap}{U^{\mathrm{a}\prime}}
\newcommand{\uav}{U^{\mathrm{a, v}}}
\newcommand{\rd}{R^{\mathrm{d}}}
\newcommand{\ra}{R^{\mathrm{a}}}
\newcommand{\pd}{P^{\mathrm{d}}}
\newcommand{\pa}{P^{\mathrm{a}}}
\newcommand{\cv}{c^{\mathrm{v}}}
\newcommand{\cw}{c^{\mathrm{w}}}

\begin{document}

\maketitle

\begin{abstract}
Community engagement plays a critical role in anti-poaching efforts, yet existing mathematical models aimed at enhancing this engagement often overlook direct participation by community members as alternative patrollers. Unlike professional rangers, community members typically lack flexibility and experience, resulting in new challenges in optimizing patrol resource allocation. To address this gap, we propose a novel game-theoretic model for community-participated patrol, where a conservation agency strategically deploys both professional rangers and community members to safeguard wildlife against a best-responding poacher. In addition to a mixed-integer linear program formulation, we introduce a Two-Dimensional Binary Search algorithm and a novel Hybrid Waterfilling algorithm to efficiently solve the game in polynomial time. 
Through extensive experiments and a detailed case study focused on a protected tiger habitat in Northeast China, we demonstrate the effectiveness of our algorithms and the practical applicability of our model.
\end{abstract}

\begin{links}
    \link{Code}{https://github.com/YvonneWu10/Community-Participated-Patrol}
\end{links}

\section{Introduction}
\label{sec:introduction}



Community engagement has become increasingly recognized as a vital element in the fight against wildlife poaching \cite{gill2014community,nubani2023community,wilson2021community}. By involving local communities directly in conservation efforts, agencies can leverage local knowledge and foster a sense of ownership over the protection of natural resources. Among various strategies, community-participated patrols, where community members actively engage in monitoring and protecting wildlife, have gained popularity as an effective means of enhancing anti-poaching initiatives \cite{masse2017inclusive,danoff2022individual}. These patrols not only supplement the efforts of professional rangers but also strengthen the ties between conservation agencies and the communities they serve.

Despite the rise of community-participated patrols, existing game-theoretic models for anti-poaching resource allocation have largely overlooked the direct role of community members as patrollers. Most models focus on indirect engagement, like community reporting poacher locations to rangers \cite{sjostedt2022governance,linkie2015editor,huang2020green,shen2020follow,shen2024extensive}, without addressing the complexities of deploying community members in the field. Differences in flexibility and experience between community members and professional rangers pose unique challenges in optimizing patrols, which remain underexplored. 
For example, in Northeast China, each community member is assigned to a 2km by 2km area and will patrol the same area at least twice a week throughout the patrol season. In contrast, professional rangers cover multiple areas every week, with the choices of areas changing weekly. 
 
To address this gap, we propose a novel game-theoretic model that explicitly incorporates community-participated patrols into the strategic allocation of anti-poaching resources. Our model enables conservation agencies to assign both professional rangers and community members (or villagers) to multiple target areas, accounting for the poacher’s choice of the target with the highest expected utility after observing the patrol strategy. Rangers follow a mixed strategy, with assignments based on a probability distribution over targets. The randomness in this strategy makes it harder for poachers to avoid detection. In contrast, villagers, due to limited flexibility, follow a deterministic strategy, with each assigned to a specific target.

To solve the game, we first present a mathematical program-based solution. The deterministic allocation for villagers introduces integer variables, resulting in an NP-hard mixed-integer linear program. We then design two polynomial-time algorithms that leverage key properties of the problem. The first, the Two-Dimensional Binary Search algorithm, is an approximation method that performs binary searches on both the number of villagers and the number of rangers assigned to the poacher's chosen target. The second, the Hybrid Waterfilling algorithm, provides an exact solution by combining the "water-filling" idea~\cite{kiekintveld2009computing} from security games with binary search and iterative target swapping for villagers.
We validate our approach's effectiveness and computational efficiency through synthetic data experiments. We further run an extensive case study on a protected tiger habitat in Northeast China with real-world data, demonstrating the potential of our model in real-world conservation scenarios.
We have presented the model to conservation agencies in 13 countries and we plan to work with the local forest bureau in Northeast China to adjust future patrol resource allocation based on the case study results.

\section{Related Work}
\label{sec:related_work}




Many works on patrol allocation in security and environmental sustainability domains consider homogeneous patrol resources \cite{fang2015security,johnson2012patrol}.  
For heterogeneous patrol resources, there have been several works studying the synergy between rangers and mobile sensors for anti-poaching \cite{basilico2017coordinating,xu2018strategic,bondi2020signal}.
The sensors can be used for detecting poaching and informing patrollers, but can not stop the poachers themselves. The sensors can also signal the poachers that rangers are coming to deter them from poaching. 
Therefore, these works pay attention to the design of the joint allocation of rangers and sensors to raise the probability that the rangers come when poaching is detected and the strategic signaling scheme of the mobile sensors to better deter poachers. Unlike sensors, villagers can stop poaching by removing the snares but can only follow deterministic allocation. Therefore, the main technical focus of our work is the combination of deterministic allocation of villagers and random allocation of rangers.

In addition, \citet{mc2016preventing} proposes a model for the optimal selection and deployment of security resource teams with varying effectiveness and costs. 
The model is extended in \citet{mccarthy2018price}, which considers the inflexibility of resources that defenders are only allowed to be distributed to a set of targets in given periods, resulting in an NP-Hard problem. In contrast, we specify the inflexibility that villagers can not be allocated to different targets and provide a polynomial-time algorithm.

Coordinating multiple defenders is widely studied in security domains \cite{mutzari2021coalition,castiglioni2021committing,gutierrez2023cooperative}. These works usually consider that different defenders have different objectives and focus on the design of stable agreements among defenders. In our case, we view the rangers and villagers as being fully cooperative.

The idea of ``water-filling'' has been used in the design of algorithms for solving mathematical models for security problems \cite{kiekintveld2009computing,nguyen2015making,nguyen2019imitative,gan2019manipulating}. 
By treating each target as a bucket and patrol resources as water, the algorithm fills water to the bucket with the minimum water level. A key factor in using the water-filling algorithm is that the patrol resources can be divided and allocated to multiple targets at will so that they can be treated as water. However, the villagers in our problem can not be distributed to different targets, which breaks this property. Extending the algorithm to handle the villagers is a highly non-trivial task, and we design our Hybrid Waterfilling Algorithm by combining water filling with binary search and iterative swapping to address the challenge as detailed in later sections.

\section{Problem Formulation}
\label{sec:problem_formulation}

We propose and study the \textit{Resources Allocation of Community Participated Patrol} (RACPP) problem. This problem formulation builds upon the Stackleberg Security Game (SSG) formulation~\cite{Tambe_2011} where a defender allocates patrol resources to protect a set of targets against a best-responding attacker. However, the key difference in RACPP is that two types of patrol resources co-exist: professional rangers and community members who are less flexible and effective. In the anti-poaching domain, community members are from nearby villages and towns who are less experienced in finding poaching tools during patrols and are often only willing to go to a fixed area (i.e., one target) for patrols throughout the patrol season as it is easier for them to plan their other daily work. 
For expository purposes, we will refer to them as villagers in the rest of the paper.

More concretely, a defender can allocate a group of $\rp$ rangers and $\rv$ villagers to protect $n$ targets $T = \{0, 1, \dots, n - 1\}$. A ranger can distribute their\footnote{We use their instead of his or her in this paper.} efforts among multiple targets, while a villager can only be allocated to a single target. We denote the defenders' defensive strategy profile as a tuple $(\b p, \b v)$, where $\b p = (p_{0}, p_{1}, \ldots, p_{n-1})$ and $\b v = (v_{0}, v_{1}, \ldots, v_{n-1})$. Here, $p_{i}\in \mathbb R_{\geq 0}$ is the amount of rangers' effort distributed to target $i$ and $v_{i}\in \mathbb N$ is the number of villagers patrolling target $i$. A \textit{valid} defender strategy profile must satisfy 
\begin{equation}
    \begin{array}{c}
        \sum_{i \in T}p_i\leq \rp, \quad \sum_{i \in T}v_i\leq \rv. \label{equation:resource}
    \end{array}
\end{equation}

The attacker will select a target $i \in T$ to attack to maximize their expected utility after observing the defender strategy. The probability of successfully attacking the target depends on the \emph{coverage} of that target provided by the defender. Specifically, let the defense effectiveness of one unit of ranger and one villager be $\ep$ and $\ev$, respectively. The total coverage on target $i$ is 
\begin{equation}
c_{i} = \min(\ep \cdot p_{i} + \ev \cdot v_{i}, 1).\label{equation:ci}
\end{equation}
We use the coverage vector $\b c = (c_{0}, c_{1}, \ldots, c_{n-1})$ to denote the coverage on all targets. If target $i$ is attacked, the probability of successfully defending the target is $c_{i}$. The attacker and defender receive rewards and penalties based on the outcome of the attack. If target $i$ is successfully defended, the defender receives reward $\rd_{i}$ and the attacker receives penalty $\pa_{i}$. Otherwise, the defender receives penalty $\pd_{i}$ and the attacker receives reward $\ra_{i}$. Here, we assume $\rd_{i} \geq 0 \geq \pd_{i}$ and $\ra_{i} \geq 0 \geq \pa_{i}$. Then, the expected utility of defenders and the attacker are respectively 
\begin{equation}
    \begin{array}{c}
        \ud_{i} = \rd_{i} \cdot c_{i} + \pd_{i} \cdot (1 - c_{i}), \\
         \ua_{i} = \ra_{i} \cdot (1 - c_{i}) + \pa_{i} \cdot c_{i}.\label{equation:pv_to_ua}
    \end{array}
\end{equation}

The attacker always chooses to attack the target that maximizes their expected utility. If there are multiple targets that maximize the attacker's expected utility, the attacker will choose the one that maximizes the defender's expected utility following the standard SSG model \cite{Tambe_2011}. Thus, given the defender strategy profile $(\b p, \b v)$, the attacker's response is fixed. We can then define the defenders' expected utility as a function of the defender strategy profile, $u(\b p, \b v)$. The defenders' goal is to maximize their expected utility by adjusting their defensive strategy $(\b p, \b v)$ as the attacker selects target $i$ to attack with \textit{best response}.

\begin{definition}
    \label{def:input_instance}
    An \textbf{input instance} of the RACPP is a tuple $\I = (n, \rp, \rv, \ep, \ev, \b{\rd}, \b{\pd}, \b{\ra}, \b{\pa})$.
\end{definition}

\section{Algorithms for RACPP}
\label{sec:solutions}

In this section, we will present our algorithms to solve the RACPP problem. First, in \cref{subsec:mixed_integer_linear_program_solution}, we show that the RACPP problem can be formulated as a mixed-integer linear program (MILP). This MILP is easy to understand, but its runtime is exponential in the worst case. We then show in \cref{subsec:two_dimensional_binary_search} our Two-Dimensional Binary Search algorithm which solves the RACPP problem to any given accuracy $\varepsilon$ in $O(n^2\log\frac{M}{\varepsilon})$, where $M$ is the maximum absolute value of the reward and penalties. Finally, in \cref{subsec:exact_algorithm}, we present an exact algorithm named Hybrid Waterfilling algorithm that solves RACPP precisely in $O(n^4\log n)$.

\subsection{Mixed-Integer Linear Program Solution}
\label{subsec:mixed_integer_linear_program_solution}

The RACPP problem is a Stackelberg game \cite{stackelberg1934marktform}, where the defender is the leader and the attacker is the follower. Suppose we already know that target $i^{*}$ is the target to be attacked in equilibrium. To ensure target $i^{*}$ is the attacker's best response, we need $\ua_{i^{*}} \geq \ua_i, \forall i \in T$ \cite{conitzer2006computing}, and we would like to maximize the defender utility $\ud_{i^{*}}$ subjected to \eqref{equation:resource}, \eqref{equation:ci} and \eqref{equation:pv_to_ua}. If we consider the defender strategy profile $(\b p, \b v)$ as a set of variables, then, the RACPP problem can be formulated as an optimization problem. This problem can be converted into a MILP by using common techniques from the MILP literature \cite{bradley1977applied} to linearize \eqref{equation:ci}. To enforce $c_{i}$ to be $\min(\ep \cdot p_{i} + \ev \cdot v_{i}, 1)$, we introduce a large number $M$ (dependent on the instance), continuous variables $\boldsymbol{\delta}$ and binary variables $\b w$. For each $i \in T$, we require $w_i \leq c_i \leq 1, \delta_i \leq M \cdot w_i$ and $c_{i} + \delta_{i} = \ep \cdot p_{i} + \ev \cdot v_{i}$. When $\ep \cdot p_{i} + \ev \cdot v_{i} \leq 1$, $w_{i}$ and $\delta_{i}$ are limited to 0, thus $c_{i} = \ep \cdot p_{i} + \ev \cdot v_{i}$; Otherwise, $c_{i}$ is set to 1 since $w_{i}$ is limited to 1. The complete MILP is as follows.

$$
\begin{array}{l|ll}
	\textbf{Maximize} & \ud_{i^{*}}\\
	\textbf{Subject to} & \ua_{i^{*}} \geq \ua_{i} & (\forall i \in T)\\
    & c_{i} + \delta_{i} = \ep \cdot p_{i} + \ev \cdot v_{i} & (\forall i \in T)\\
    &  \delta_{i} \leq M \cdot w_{i}, \quad \quad w_{i} \leq c_{i} \leq 1 & (\forall i \in T) \\
    & \ua_{i} = \ra_{i} \cdot (1 - c_{i}) + \pa_{i} \cdot c_{i} & (\forall i \in T) \\
    & \ud_{i} = \rd_{i} \cdot c_{i} + \pd_{i} \cdot (1 - c_{i}) & (\forall i \in T) \\
    & p_{i} \geq 0, \quad \quad \quad \quad\  v_{i} \in \mathbb{N} & (\forall i \in T) \\
    & \delta_{i} \geq 0, \quad \quad \quad \quad\ w_{i} \in \{0, 1\} & (\forall i \in T) \\
	& \sum_{i \in T} p_{i} \leq \rp,\quad \sum_{i \in T} v_{i} \leq \rv \\
\end{array}
$$

By enumerating all targets as the attacked target $i^{*}$, we can solve the RACPP problem by solving the MILP for each target. However, the runtime of the MILP solution is exponential in the worst case since MILP is NP-hard \cite{karp2010reducibility}. In the rest of the paper, we aim to design a polynomial-time algorithm to solve the RACPP problem.

\subsection{A Polynomial Approximate Algorithm: Two-Dimensional Binary Search}
\label{subsec:two_dimensional_binary_search}

In this section, we will present a polynomial-time algorithm that solves the RACPP problem to any desired accuracy $\varepsilon$. The algorithm is based on a two-dimensional binary search.

To begin with, we first consider the following decision problem: given a target $i^{*}$ and a fixed defender strategy $(p_{i^{*}}, v_{i^{*}})$ on $i^{*}$, can we find a strategy profile $(\b p, \b v)$ such that the attacker will choose to attack the target $i^{*}$ with best response? We can solve this problem by greedily distributing the remaining resources to other targets. The following \cref{alg:binary_judge} checks whether a consistent strategy exists.

\begin{algorithm}[htbp]
    \caption{Checking whether a consistent strategy exists}
    \label{alg:binary_judge}

    \textbf{Input}: Input instance $\I$, target $i^{*}$, strategy $(p_{i^{*}}$, $v_{i^{*}})$ on $i^{*}$

    \textbf{Output}: Whether a consistent strategy $(\b p, \b v)$ exists

    \begin{algorithmic}[1] 
        \STATE Compute $\ua_{i^{*}}$ using \eqref{equation:ci}, \eqref{equation:pv_to_ua} with $(p_{i^{*}},v_{i^{*}})$.
        \IF{$\exists i\in T, \ua_{i^{*}} < \pa_{i}$}
            \RETURN $\mathrm{False}$.
        \ENDIF
        \STATE Let $(\rp_{\mathrm{remain}}, \rv_{\mathrm{remain}}) \gets (\rp - p_{i^{*}}, \rv - v_{i^{*}})$.
        \STATE Let $\delta_{i} \gets 0$ \textbf{for all} $i \in T$.
        \FOR{$i \in \{0,1,\dots,n-1\} \setminus \{i^{*}\}$}
            \STATE Let $c_{\mathrm{min}, i} \gets \text{the minimum $c_i$ to ensure $\ua_{i} \leq \ua_{i^{*}}$}$.
            \STATE Let $v_{\mathrm{cnt}, i} \gets \min \left( \lfloor c_{\mathrm{min}, i}/\ev\rfloor, \rv_{\mathrm{remain}} \right)$.
            \STATE Let $\rv_{\mathrm{remain}} \gets \rv_{\mathrm{remain}} - v_{\mathrm{cnt}, i}$.
            \STATE Let $\delta_i \gets c_{\mathrm{min}, i} - v_{\mathrm{cnt}, i}\cdot \ev$.
        \ENDFOR
        \FOR{$i \in \{0, 1, \dots, \min(n, \rv_{\mathrm{remain}})-1\}$}
            \STATE Let $\text{the largest $\delta_j\ (j\in T)$} \gets 0$.
        \ENDFOR
        \RETURN $\sum_{i \neq i^{*}} \delta_i \leq \rp_{\mathrm{remain}} \cdot \ep$.
    \end{algorithmic}
\end{algorithm}

\begin{definition}
    \label{def:min_coverage}
    Given an utility $u$, the \textbf{minimum valid coverage} is a vector $\b c_{\mr{min}}(u)$, where $c_{\mr{min}, i}(u)$ is the minimum coverage on target $i$ such that $\ua_{i} \leq u$.
\end{definition}

\begin{definition}
\label{def:wasted_villager_coverage}
    Given a villager strategy $\b v$ and an utility $u$, the \textbf{wasted villager coverage} is a vector $\b{\cw}(\b v, u)$, where $\cw_{i}(\b v, u) = \max(v_{i} \cdot \ev - c_{\mathrm{min}, i}(u), 0)$, and the total wasted villager coverage is $\mr{scw}(\b v, u) = \sum_{i \neq i^{*}}\cw_{i}(\b v, u)$.
\end{definition}

\cref{alg:binary_judge} first checks whether there are targets that allow the attacker utility to be higher than $\ua_{i^{*}}$ no matter how many resources are allocated to them (Lines 2). If there is such a target, then it is impossible to make $i^{*}$ the attacker's best response (Line 3). Otherwise, the algorithm will try to allocate resources to other targets to ensure that the attacker's utility is no more than $\ua_{i^{*}}$ (Lines 4 to 12). To do this, we first calculate the minimum coverage $c_{\mathrm{min}, i}$ for each target $i$ (Line 7). The problem is then to check if it is feasible to achieve the required minimum coverage on each target. 

We allocate the resources greedily: we try first to allocate villagers and then rangers. Intuitively, this is because villagers can only be allocated to one target as a whole, while rangers are more flexible in distributing their effort to multiple targets. Specifically, we first try to allocate as many villagers as possible so that the coverage they provide is fully utilized (Lines 8 to 10). If some villagers remain, it means that allocating them would cause some targets to have more coverage than necessary. We then allocate these villagers to the targets that minimize the wasted coverage (Lines 11 to 12). Finally, we check whether there are enough ranger efforts to cover the remaining needs on all targets (Line 13). \cref{alg:binary_judge} works in $O(n)$ time to check whether there is a consistent strategy. Formally, we have the following lemma.

\begin{restatable}{lemma}{BinaryJudgeCorrectness}
    \label{lemma:binary_judge_correctness}
    
    \cref{alg:binary_judge} returns $\mathrm{True}$ if and only if there exists a valid defender strategy profile $(\mathbf p, \mathbf v)$ such that $p_{i^{*}} = p, v_{i^{*}} = v$ and $i^{*}$ is the attacker's best response.
\end{restatable}

The proof of \cref{lemma:binary_judge_correctness} is deferred to \cref{appsub:proof_of_lemma_binary_judge}. 

Now that we have the ability to judge whether a consistent strategy exists, for target $i^{*}$ and fixed defender strategy $(p_{i^{*}}, v_{i^{*}})$ on $i^{*}$. For those pairs of $(p_{i^{*}}, v_{i^{*}})$ with a consistent strategy, we would like to find the one that maximizes the coverage $c_{i^{*}} = \min(p_{i^{*}}\cdot \ep + v_{i^{*}}\cdot \ev,1)$ on $i^{*}$. We will use a two-dimensional binary search to find this maximum coverage. To do this efficiently, we need to establish two monotonicity lemmas: \cref{lemma:mono_patroller_and_villager,lemma:mono_coverage_unchanged}.

\begin{restatable}{lemma}{MonoPatrollerAndVillager}
    \label{lemma:mono_patroller_and_villager}
    
    Let $i^{*}$ be a target, and let $(\b p, \b v)$ be a valid defender strategy profile such that $i^{*}$ is the attacker's best response. Then, $\forall 0 \leq p\leq p_{i^{*}}$, $0 \leq v\leq v_{i^{*}}$ $(v \in \mathbb{N})$, there is a valid defender strategy profile $(\b p', \b v')$ such that $p'_{i^{*}}=p, v'_{i^{*}}=v$ and $i^{*}$ is still one of attacker's best responses.
\end{restatable}

We present the proof of \cref{lemma:mono_patroller_and_villager} in \cref{appsub:proof_of_lemma_mono_patroller_and_villager}. Intuitively, \cref{lemma:mono_patroller_and_villager} shows for a target $i^{*}$ that an attacker will attack with the best response under some defender strategy $(\b p, \b v)$, we can always reduce the resources allocated to $i^{*}$ while keeping the attacker's best response unchanged. 

We then move on to another monotonicity lemma. Rangers and villagers can substitute each other on a specific target provided that they are of the same effectiveness, e.g., one villager can be replaced with $\ev / \ep$ units of ranger effort. However, since a villager can only patrol on one target while a ranger can distribute their efforts among multiple targets, ranger efforts are more flexible, and thus more useful when the two kinds of resources can be converted into the same amount of coverage. Therefore, converting ranger efforts to villagers on target $i^{*}$ makes it easier to satisfy the coverage needs of the other targets to make $i^{*}$ the attacker's best response. We formally state this intuition as follows.

\begin{restatable}{lemma}{MonoCoverageUnchanged}
    \label{lemma:mono_coverage_unchanged}
    
    Let $i^{*}$ be a target, and let $(\b p,\b v)$ be a valid defender strategy profile such that $i^{*}$ is the attacker's best response. Then, for any $p\in \mathbb R_{\geq 0},v \in \mathbb{N}$ such that $p \cdot \ep + v \cdot \ev = p_{i^{*}} \cdot \ep + v_{i^{*}} \cdot \ev$, $p \leq p_{i^{*}}$ and $v \geq v_{i^{*}}$, there exists a valid defender strategy profile $(\mathbf p',\mathbf v')$ such that $p'_{i^{*}}=p, v'_{i^{*}}=v$ and $i^{*}$ is still one of attacker's best responses.
\end{restatable}

The proof of \cref{lemma:mono_coverage_unchanged} is deferred to \cref{appsub:proof_of_lemma_mono_coverage_unchanged}.

With the lemmas above, we are ready for our algorithm to compute the maximum defender utility in the RACPP problem approximately. The algorithm is shown in \cref{alg:binary}.

\begin{algorithm}[htbp]
    \caption{Two-dimensional binary search}
    \label{alg:binary}

    \textbf{Input}: Input instance $\I$ and precision $\varepsilon$

    \textbf{Output}: Approximate maximum defender utility

    \begin{algorithmic}[1]
        \STATE Let $u_{\mathrm{ans}} \gets -\infty$.
        \FOR{$i^{*} \in \{0, 1, \dots, n-1\}$}
            \IF{\cref{alg:binary_judge} returns $\mathrm{False}$ on $(\I, i^{*}, 0, 0)$}
                \STATE \textbf{continue}
            \ENDIF
            \STATE Let $(v_{\mathrm{left}},v_{\mathrm{right}}) \gets (0,\rv)$.
            \WHILE{$v_{\mathrm{left}} \leq v_{\mathrm{right}}$}
                \STATE Let $v_{\mathrm{cur}} \gets \lfloor(v_{\mathrm{left}} + v_{\mathrm{right}})/{2}\rfloor$.
                 \IF{\cref{alg:binary_judge} returns $\mathrm{True}$ on $(\I, i^{*}, 0, v_{\mathrm{cur}})$}
                    \STATE Let $v_{\mathrm{left}} \gets v_{\mathrm{cur}} + 1$.
                    \STATE Let $v_{i^{*}} \gets v_{\mathrm{cur}}$.
                \ELSE
                    \STATE Let $v_{\mathrm{right}} \gets v_{\mathrm{cur}} - 1$.
                \ENDIF
            \ENDWHILE
            \STATE Let $(p_{\mathrm{left}},p_{\mathrm{right}}) \gets (0,\rp)$.
            \WHILE{$p_{\mathrm{right}} - p_{\mathrm{left}} > \varepsilon$}
                \STATE Let $p_{\mathrm{cur}} \gets (p_{\mathrm{left}} + p_{\mathrm{right}})/{2}$.
                \IF{\cref{alg:binary_judge} returns $\mathrm{True}$ on $(\I, i^{*}, p_{\mathrm{cur}}, v_{i^{*}})$}
                    \STATE Let $p_{\mathrm{left}} \gets p_{\mathrm{cur}}$.
                    \STATE Let $p_{i^{*}} \gets p_{\mathrm{cur}}$.
                \ELSE
                    \STATE Let $p_{\mathrm{right}} \gets p_{\mathrm{cur}}$.
                \ENDIF
            \ENDWHILE
            \STATE Compute $\ud_{i^{*}}$ using $(p_{i^{*}}, v_{i^{*}})$.
            \STATE Let $u_{\mathrm{ans}} \gets \max(u_{\mathrm{ans}}, \ud_{i^{*}})$.
        \ENDFOR
        \RETURN $u_{\mathrm{ans}}$.
    \end{algorithmic}
\end{algorithm}

\cref{alg:binary} enumerates all possible targets $i^{*}$ as the attacker's best response (Line 2) and tries to find the maximum coverage on $i^{*}$ as mentioned above. To do this, it first finds the maximum number of villagers $v_{i^{*}}$ that can be deployed on target $i^{*}$ such that we can still make target $i^{*}$ the attacker's best response (Lines 5 to 12). Using the monotonicity established in \cref{lemma:mono_patroller_and_villager} as well as \cref{alg:binary_judge}, we can use binary search to find such $v_{i^{*}}$. Note that as we maximize the villagers used on target $i^*$, we also maximally substitute any ranger efforts that will be allocated to $i^*$ with villagers. According to \cref{lemma:mono_coverage_unchanged}, this will not make it harder to satisfy the coverage needs of the other targets.

We then fix the number of villagers deployed at $i^{*}$ and find the maximum possible ranger efforts $p_{i^{*}}$ on target $i^{*}$ (Lines 13 to 20). Again by \cref{lemma:mono_patroller_and_villager}, $p_{i^{*}}$ can also be found using binary search. This pair of $(p_{i^{*}}, v_{i^{*}})$ is guaranteed to be optimal because we first try to maximize $v_{i^{*}}$, and then $p_{i^{*}}$. Finally, after enumerating each possible $i^{*} \in T$, the algorithm will return the best defender utility (Lines 21 to 22). Throughout the process, \cref{alg:binary} makes a total of $O(n\log \frac{M}{\varepsilon})$ calls to \cref{alg:binary_judge}, where $M$ is the maximum absolute value of the input variables. Therefore, \cref{alg:binary} works in $O(n^{2}\log \frac{M}{\varepsilon})$ time.

\begin{restatable}{theorem}{BinaryBound}
    \label{theorem:binary_bound}
    Let the absolute values of the input variables be bounded by $M$. \cref{alg:binary} generates a valid defender strategy profile $(\b p, \b v)$ in $O(n^{2} \cdot \log\frac{M}{\varepsilon})$ time such that for the optimal defender strategy profile $(\b p^*, \b v^*)$,
    $$
    u(\b p^*, \b v^*) - u(\b p, \b v) < \ep \cdot 2M\varepsilon.
    $$
\end{restatable}

The proof of \cref{theorem:binary_bound} is presented in \cref{appsub:proof_of_theorem_binary_bound}.

\subsection{A Polynomial Exact Algorithm: Hybrid Waterfilling Algorithm}
\label{subsec:exact_algorithm}

The Two-Dimensional Binary Search algorithm can solve the RACPP problem to any given accuracy $\varepsilon$. However, it is not able to solve the problem precisely. Therefore, in this section, we study exact polynomial-time algorithms for the RACPP problem. We will present a polynomial-time algorithm, the Hybrid Waterfilling algorithm, that solves the RACPP problem precisely.

\begin{figure*}[t]
	\centering
	\begin{subfigure}{0.235\textwidth}
	  \centering
	  \includegraphics[width=\linewidth]{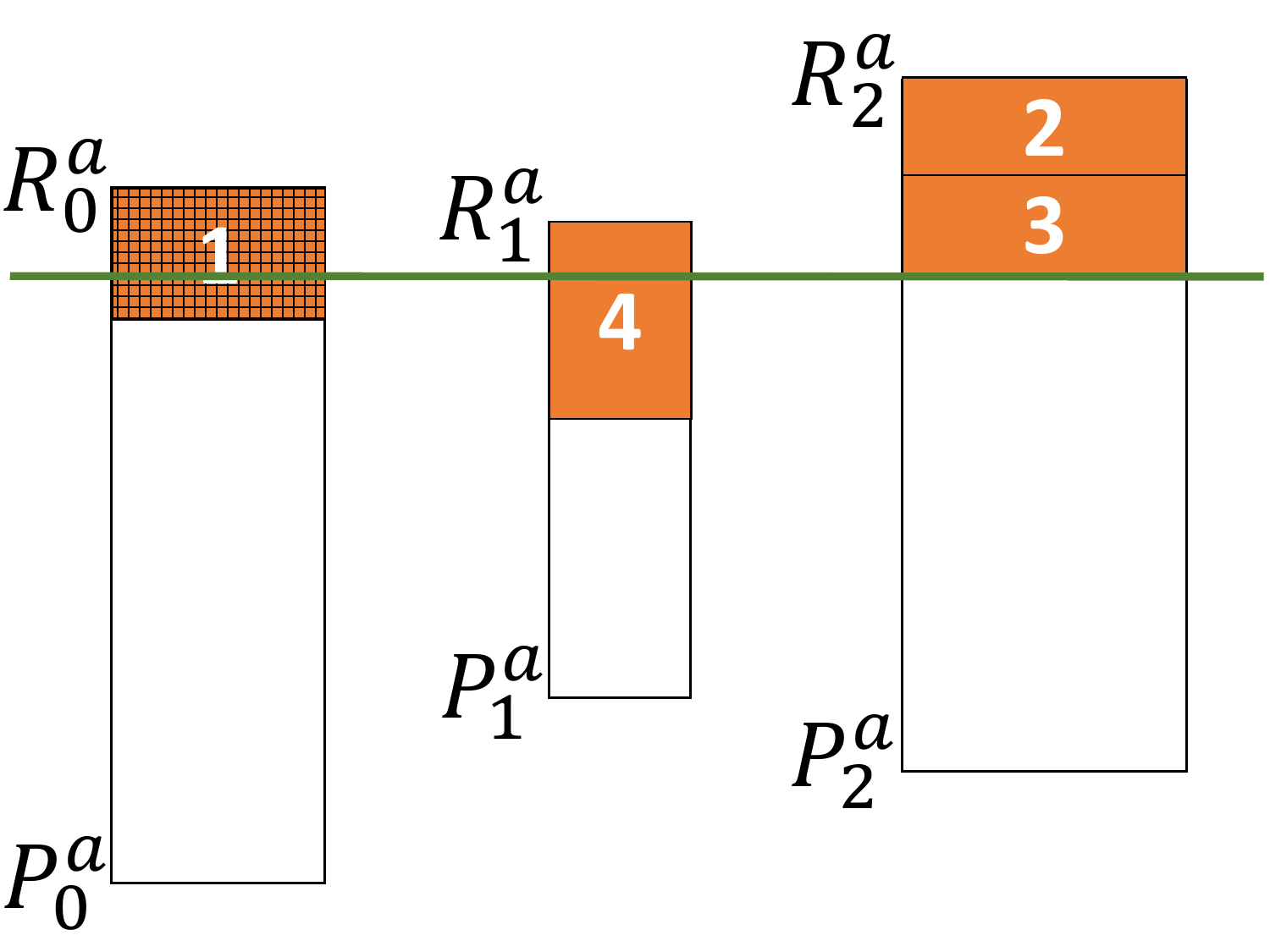}
      \caption{Greedy for villagers}
	  \label{subfigure:step2}
	\end{subfigure}
	\hfill
	\begin{subfigure}{0.235\textwidth}
	  \centering
	  \includegraphics[width=\linewidth]{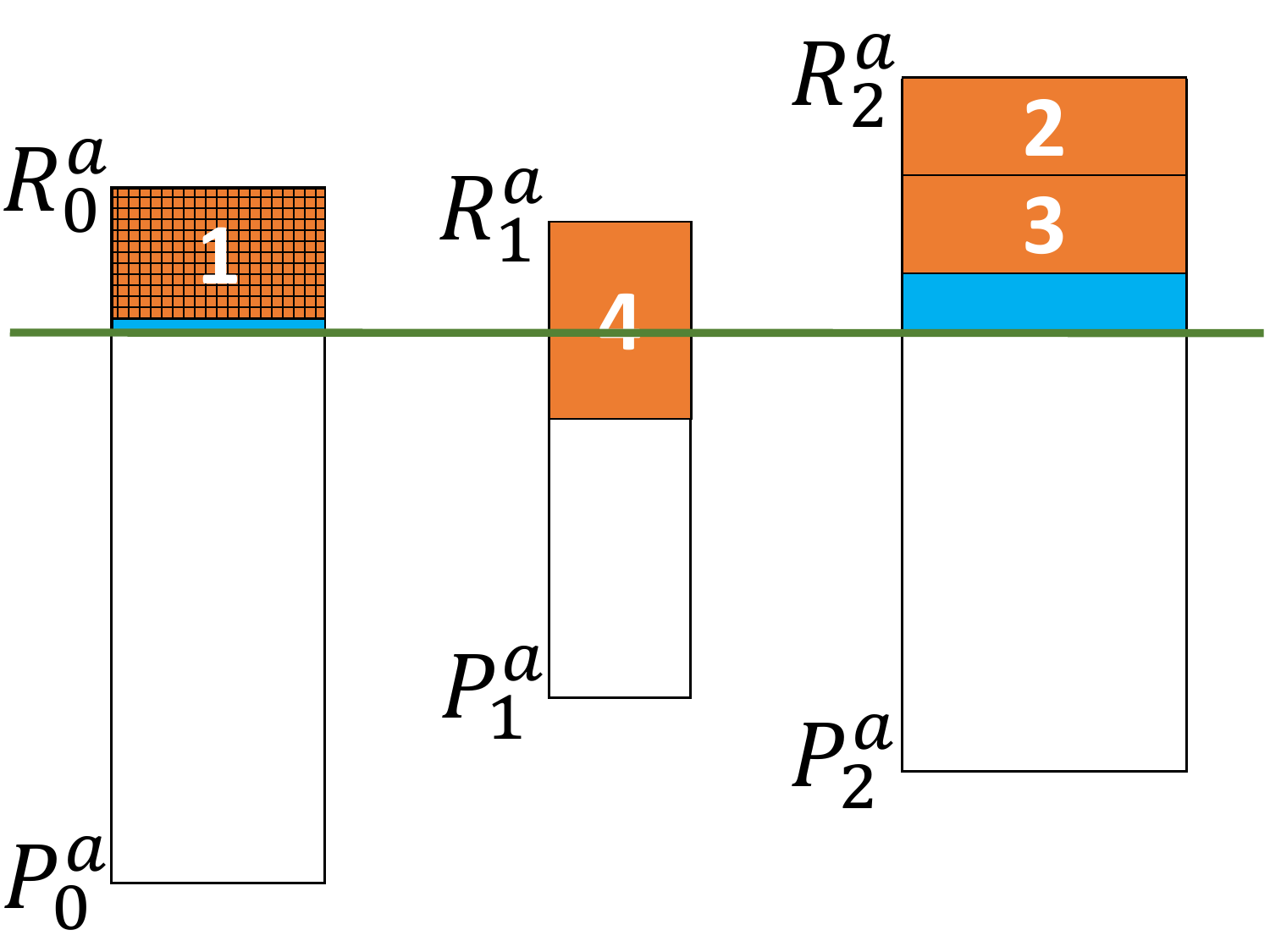}
	  \caption{Waterfilling to a critical point}
	  \label{subfigure:step3}
	\end{subfigure}
    \hfill
    \begin{subfigure}{0.235\textwidth}
        \centering
        \includegraphics[width=\linewidth]{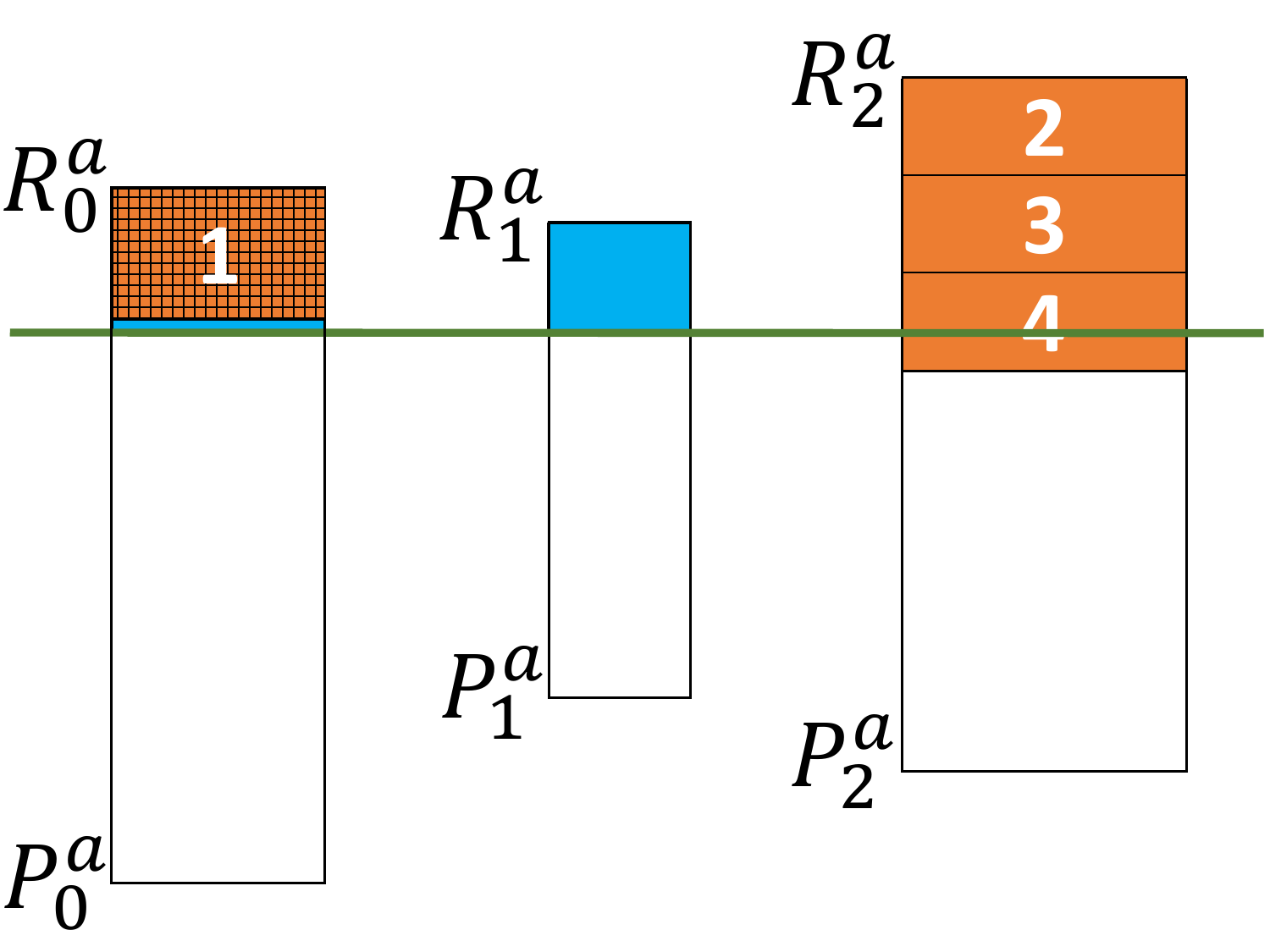}
        \caption{Trigger a \textbf{swap}}
        \label{subfigure:step4}
      \end{subfigure}
      \hfill
      \begin{subfigure}{0.235\textwidth}
        \centering
        \includegraphics[width=\linewidth]{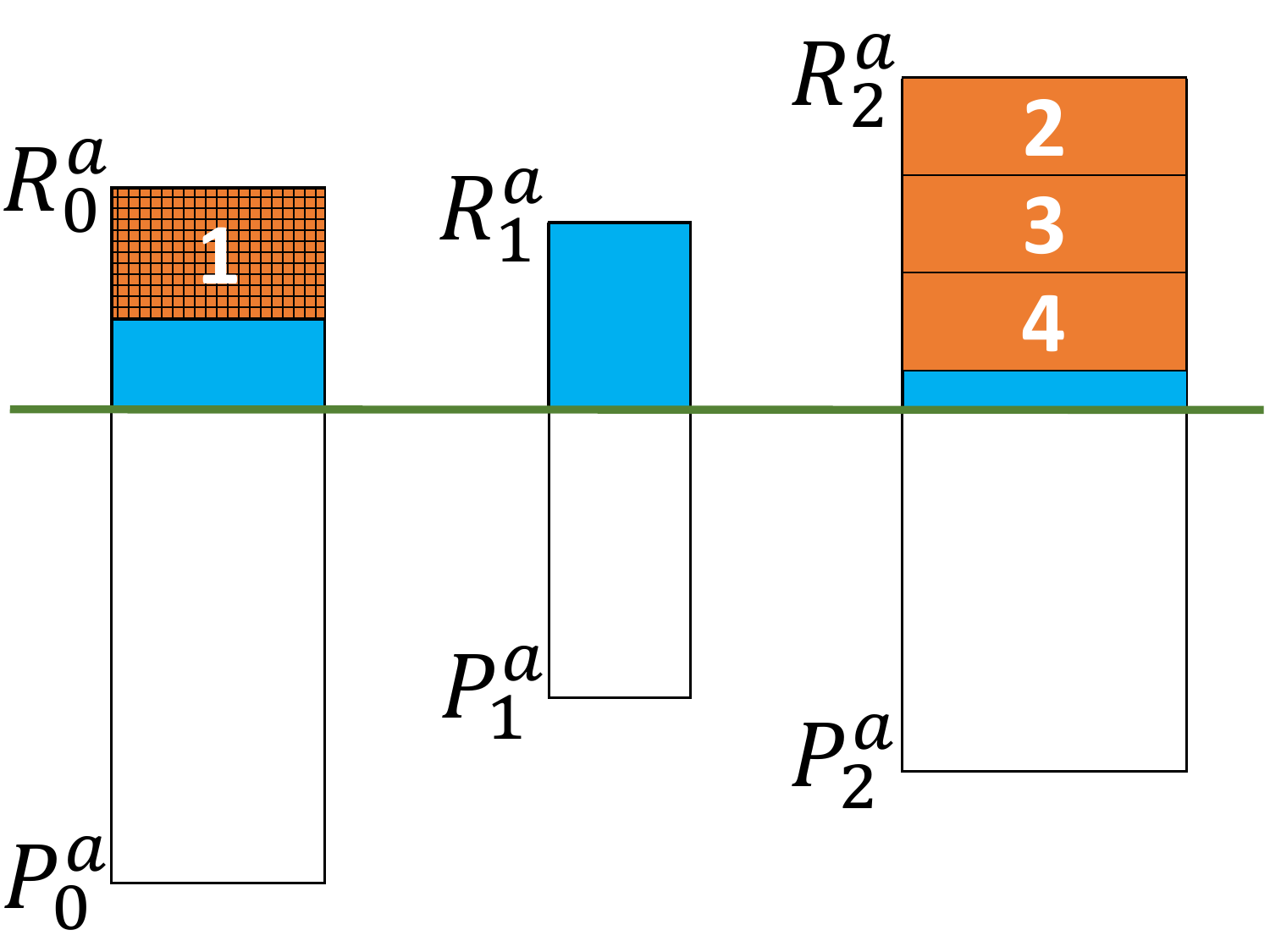}
        \caption{Finish Waterfilling}
        \label{subfigure:step5}
      \end{subfigure}

	\caption{Example of the process in \cref{alg:exact_patroller}. The 3 black rectangles represent 3 targets. The orange parts are the villager coverage and the blue ones are the ranger coverage. The green line is the utility sea level. (a) Target $i^{*} = 0$ and villager strategy $v_{i^{*}} = 1$ are given. We greedily allocate the remaining 3 villagers to target $i \in \{1, 2\}$ with the maximum $\ua_{i}$ in the order of target 2, 2, 1. (b) Critical set is $\{1\}$. We distribute ranger efforts by Waterfilling, lowering the sea level, until we reach the critical point when the area of the blue part on target 3 is equal to the area of villager 3 above the sea level. (c) We swap ranger efforts on target 2 and villager 3 on target 2. (d) We proceed with Waterfilling until rangers are used up.}
	\label{fig:exact_algorithm}
\end{figure*}

The sense of waterfilling has been applied to mathematical models in security domains \cite{kiekintveld2009computing,nguyen2015making}. The main challenge caused by the involvement of villagers is deterministic allocation. The efforts of a villager can only be allocated to one target and can not be distributed to multiple targets, which breaks the continuous change of water level in standard waterfilling algorithm \cite{thomas2006elements}.

When there are only rangers, the RACPP problem becomes a standard security game problem, which can be exactly solved by Waterfilling. Whereas when there are only villagers, the RACPP problem can also be solved exactly by greedily allocating each villager to the target with the maximum attacker utility. Our Hybrid Waterfilling algorithm combines these two methods to solve the RACPP problem precisely. We formally state the algorithm as \cref{alg:exact} in \cref{app:details_exact}. Below, we present the intuition.

\cref{alg:exact} relies on an important subproblem: given a target $i^*$ that the attacker will attack and a fixed villager strategy $v_{i^*}$ on $i^*$, how to allocate the remaining villagers and rangers to maximize the defender utility? We use a hybrid method of Greedy and Waterfilling to solve this subproblem. See \cref{fig:exact_algorithm} for an illustration.

First, we focus only on the villagers, greedily allocating them to targets $i \in T \setminus \{i^{*}\}$ with the maximum attacker utility $\ua_{i}$ (\cref{subfigure:step2}). At this time, there is a set of targets that maximize the attacker's utility. We call this set the \textbf{critical set}, and the corresponding utility the \textbf{utility sea level}. We then distribute ranger efforts in a way similar to Waterfilling, i.e., always allocating them to the critical set to lower the sea level. At some point in this process, it might be possible to exchange a villager and some ranger efforts on two targets, such that the sea level remains unchanged but the villager goes to a target with a larger \textbf{width} ($w_{i} = 1 / (\ra_{i} - \pa_{i})$) (\cref{subfigure:step3}), we call this a \textbf{critical point}, which triggers a \textbf{swap} (\cref{subfigure:step4}). After the swap, the total width of the critical set decreases, and thus the amount of ranger efforts required to lower a unit of the sea level decreases. We continue this process until all ranger efforts are used up (\cref{subfigure:step5}). We formally summarize the process as \cref{alg:exact_patroller} in \cref{app:details_exact}. The algorithm works in $O(n^{3} \log n)$ time. 

With the subproblem solved, we can then solve RACPP. To do this, we first enumerate all $i^{*} \in T$. For each given $i^{*}$, we first use binary search to find the maximum number of villagers that can be allocated to target $i^{*}$ while ensuring $i^*$ is still the attacker's best response. This is similar to the Two-Dimensional Binary Search algorithm. We then call \cref{alg:exact_patroller} to find the optimal defender utility for the given $i^{*}$ and $v_{i^{*}}$. Finally, we return the maximum defender utility among all $i^{*}$. The optimality and time complexity of the algorithm is demonstrated in the following theorem.

\begin{restatable}{theorem}{ExactCorrectness}
    \label{theorem:exact_correctness}
    
    \cref{alg:exact} generates an optimal valid defender strategy profile $(\b p, \b v)$ in $O(n^{4}\log n)$ time.
\end{restatable}

The proof of \cref{theorem:exact_correctness} is deferred to \cref{app:details_exact}. Below, we provide a proof sketch of the theorem.

\begin{proof}[Proof Sketch of \cref{theorem:exact_correctness}]
    To show the correctness of the algorithm, we first need to show its correctness for the subproblem, i.e., \cref{alg:exact_patroller} generates an optimal valid defender strategy profile $(\b p, \b v)$ for a given target $i^{*}$ and villager deployment strategy $v_{i^{*}}$. Note that in \cref{alg:exact_patroller}, the sea level $u_{\mr{cur}}$ lowers continuously as we distribute ranger efforts. For a fixed sea level $u_{\mr{cur}}$ in the Waterfilling process, \cref{alg:exact_patroller} generates a strategy profile that ensures that $i^*$ is the attacker's best responses with utility $u_{\mr{cur}}$. In this strategy profile, we define the \textbf{wasted villager coverage} $\mr{scw}(u_{\mr{cur}})$ as the coverage provided by villagers that are not needed to ensure that $i^*$ is the attacker's best response. 

    We will show that the wasted villager coverage is minimized at each sea level $u_{\mr{cur}}$, which effectively means that the amount of ranger efforts required to reach this sea level is minimized. This is done in two steps: \textbf{(i).} We show that given sea level $u_{\mr{cur}}$, we cannot redistribute the ranger efforts and at most one villager to reduce the wasted villager coverage. \textbf{(ii).} We show that the wasted villager coverage is minimized at each sea level $u_{\mr{cur}}$. For step (i), the calculation of the critical points in \cref{alg:exact_patroller} ensures it. For step (ii), we show by an adjustment argument that if (i) holds, then the structure of the wasted villager coverage ensures that redistributing more villagers cannot reduce the wasted villager coverage, either. This completes the proof of the correctness of \cref{alg:exact_patroller}. The correctness of \cref{alg:exact} then follows from the correctness of \cref{alg:exact_patroller} and the correctness of the binary search part, which is similar to \cref{alg:binary}.

    For the complexity, since \cref{alg:binary_judge} works in $O(n)$, the binary search part of \cref{alg:exact} works in $O(n^{2}\log M)$ time, where $M$ is the maximum absolute value of the input variables. For the procedure of the subproblem in \cref{alg:exact_patroller}, note that each swap operation causes a villager to be moved to a target with a larger width, and only $O(n)$ villagers can possibly be moved. Therefore, the number of swaps, i.e., the number of iterations of \cref{alg:exact_patroller} is $O(n^{2})$. Using a priority queue to simulate the procedure, each iteration takes $O(n\log n)$ time. Therefore, the total time complexity of \cref{alg:exact_patroller} is $O(n^{3}\log n)$. Assuming $M$ is polynomially bounded by $n$, the total time complexity of \cref{alg:exact} is $O(n^{4}\log n)$. This concludes the proof sketch.
\end{proof}

\section{Extensions for Practical Constraints}
\label{sec:variants}

In practice, the actual defense effectiveness varies with real-world factors. For instance, geographical features like vegetation and slope can affect the defense effectiveness, which results in differences in defense effectiveness on different targets. Moreover, individual factors, like the domain knowledge and experience of the person, can also influence defense effectiveness. Due to their lack of training, these practical factors have a greater impact on the villagers. Therefore, in this section, we consider two generalized versions of the RACPP problem, where the villagers' defense effectiveness varies with the targets and the villagers, respectively. Interestingly, for the former, our algorithms in \cref{sec:solutions} can be adapted to solve the problem exactly in polynomial time. However, in the latter case, the problem becomes NP-hard.

\paragraph{RACPP with Target-Specific Effectiveness.} We first consider the case where the defense effectiveness of villagers varies with the targets. Specifically, we redefine villagers' defense effectiveness as $\b \ev$, where $\ev_{i}$ represents the effectiveness of one villager on target $i$.

Both algorithms in \cref{sec:solutions} can be naturally adapted to solve RACPP with target-specific effectiveness. For simplicity, we only present the adapted version of \cref{alg:binary} here.

Recall that \cref{alg:binary} is based on a two-dimensional binary search that needs to check whether a consistent strategy exists for a given target $i^{*}$ and fixed defender strategy $(p_{i^{*}}, v_{i^{*}})$ on $i^{*}$ (\cref{alg:binary_judge}). In this new setting, we modify this procedure to \cref{alg:binary_judge_refined} stated in \cref{app:missing_alg_variants}. 

Like \cref{alg:binary_judge}, the general idea of \cref{alg:binary_judge_refined} is to greedily distribute the resources to other targets to ensure that the attacker's utility is no more than $\ua_{i^{*}}$. The main difference is that the coverage generated by villagers on each target is now different. For each villager, the algorithm greedily allocates them to the target, which maximizes the coverage each villager can cover. Finally, the algorithm checks whether there are enough ranger efforts as in \cref{alg:binary_judge}. The whole process can be implemented in $O(n)$ time. Formally, we have \cref{lemma:binary_judge_correctness_refined}.

\begin{restatable}{lemma}{BinaryJudgeCorrectnessRefined}
    \label{lemma:binary_judge_correctness_refined}
    
    \cref{alg:binary_judge_refined} returns $\mathrm{True}$ if and only if there exists a valid defender strategy profile $(\mathbf p, \mathbf v)$ such that $p_{i^{*}} = p, v_{i^{*}} = v$ and $i^{*}$ is the attacker's best response.
\end{restatable}

The proof of \cref{lemma:binary_judge_correctness_refined} is deferred to \cref{appsub:proof_of_lemma_binary_judge_refined}. \cref{alg:binary} in TDBS Algorithm can stay unchanged in the new setting, and the time complexity is still $O(n^{2}\log\frac{M}{\varepsilon})$.

\paragraph{RACPP with Villager-Specific Effectiveness.} 
We then consider the case where the defense effectiveness of villagers varies with the villagers. Let $V =\{0, 1, \cdots ,\rv-1\}$ be the set of villagers. We redefine villagers' defense effectiveness as a vector $\b \ev$, where $\ev_{j}$ represents the effectiveness of villager $j \in V$ on all targets.

Interestingly, the problem becomes NP-hard in this setting. We prove that RACPP with villager-specific effectiveness is NP-hard by reducing the partition problem \cite{hayes2002computing} to it. The details are deferred to \cref{appsub:proof_of_theorem_NPHardness}.

\begin{restatable}{theorem}{NPHardness}
\label{theorem:NP_hardness}
Computing the maximum defender utility and the optimal valid strategy profile of the RACPP problem with villager-specific effectiveness is NP-hard.
\end{restatable}

\section{Experiments}
\label{sec:experiments}

In this section, we conduct numerical experiments on synthetic data to evaluate the performance of our algorithms in practice. We implement our Two-Dimensional Binary Search algorithm (TDBS) and Hybrid Waterfilling algorithm (HW), along with a benchmark algorithm using Gurobi's \cite{gurobi} Mixed Integer Linear Programming (MILP) solver. Since all of the implemented algorithms have guaranteed solution quality, we observe similar performance in terms of the defender's utility from all of them. Therefore, we focus on the runtime of these algorithms. The experiments are conducted on a server with an Intel Xeon E5-2683 v4 CPU and 269.5GB RAM.

\label{subsec:performance}

\begin{figure*}[t]
    \centering
    
    \begin{subfigure}{0.48\textwidth}
        \centering
        \includegraphics[width=\linewidth]{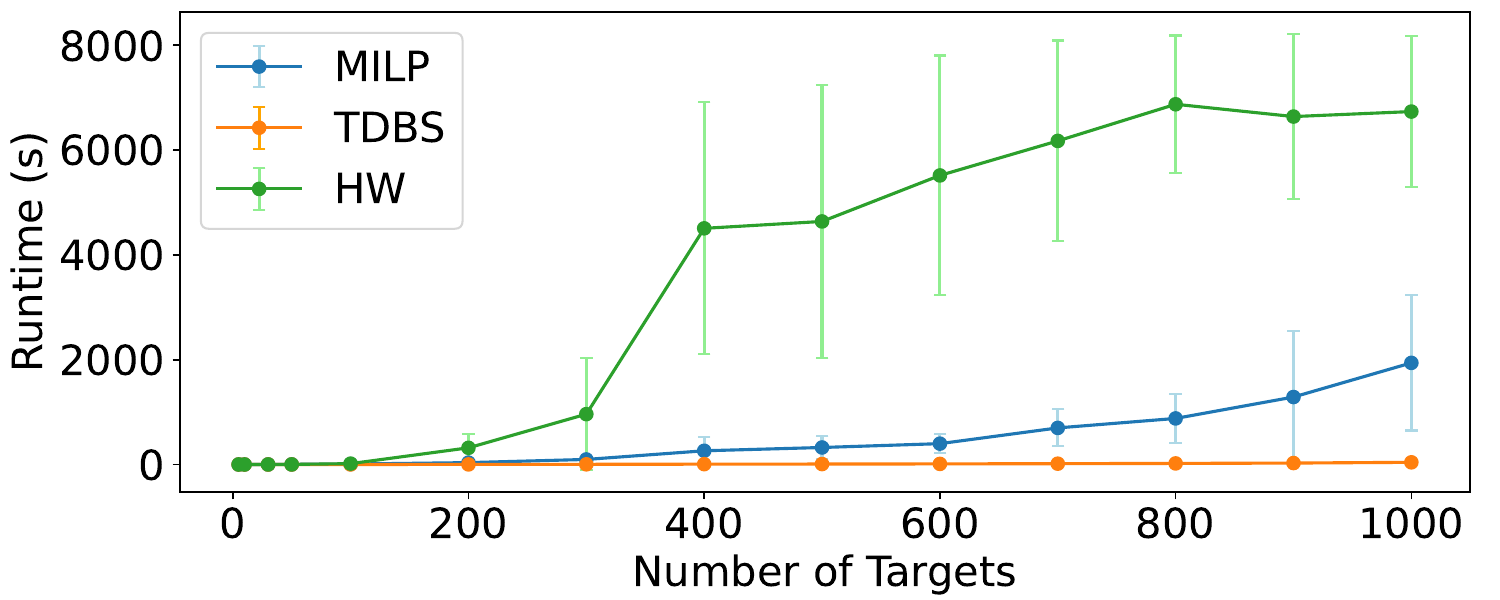}
        \caption{runtime with different $n$ and $\rp = \rv = \lfloor \frac{n}{2} \rfloor$}
        \label{subfigure:n-runtime}
    \end{subfigure}
    \hfill
    \begin{subfigure}{0.48\textwidth}
        \centering
        \includegraphics[width=\linewidth]{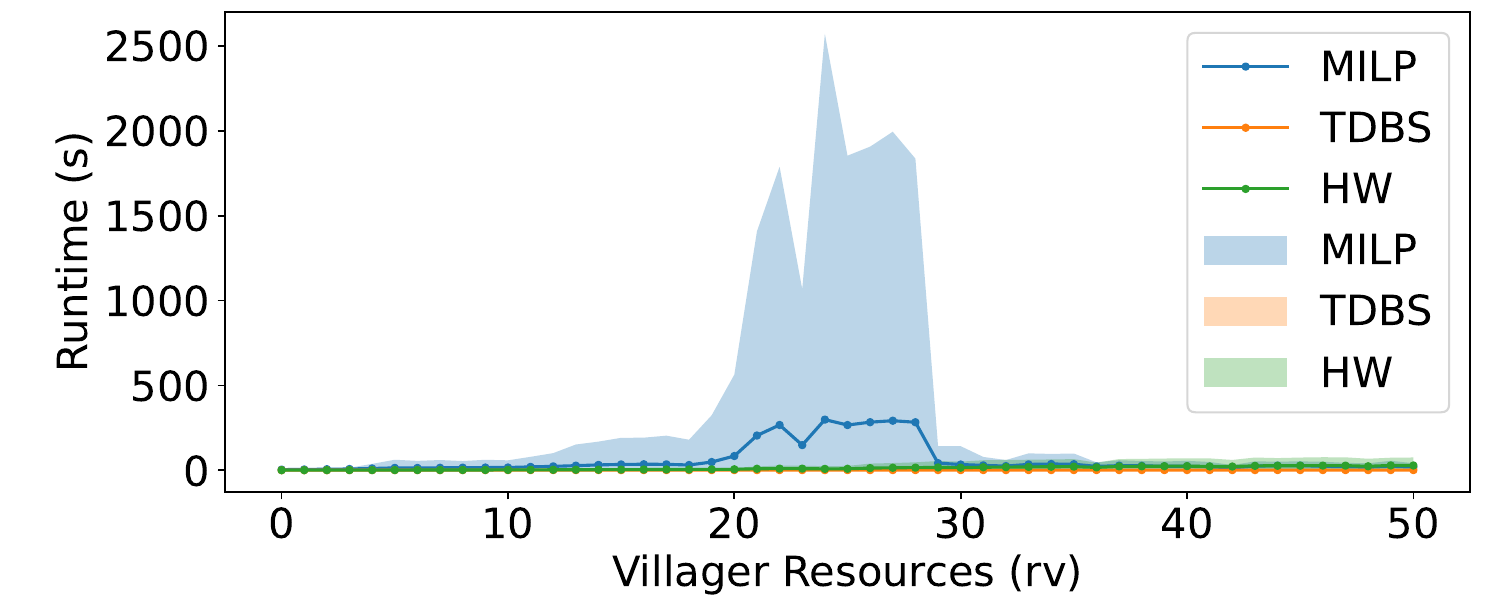}
        \caption{runtime with $n = 100$ and different $\rp,\rv$}
        \label{subfigure:rv-runtime}
    \end{subfigure}
    
    \caption{Average runtime of MILP, TDBS, and HW over 30 runs under different combinations of $(n,\rp,\rv)$. The error bars indicate the standard deviation. The shaded areas represent the ranges from the minimum to the 97th percentile. We limit the maximum runtime for MILP in \cref{subfigure:rv-runtime} to 7200 seconds. 
    }
    \label{fig:performance}
\end{figure*}

\paragraph{Experiment setup.} We evaluate the algorithms with different combinations of $(n,\rp,\rv)$. For each combination, we randomly generate $\ep$, $\ev$, $\b{\rd}$, $\b{\pd}$, $\b{\ra}$, $\b{\pa}$ with $0 < ev < ep < 1$, $\rd_{i}, \ra_{i} \in [0, 10)$ and $\pd_{i}, \pa_{i} \in [-10, 0)$ and record the runtime of the algorithms. We report the mean and the standard deviation of the runtime over 30 runs for each combination. The precision of TDBS is set to $10^{-3}$. If a single run of an algorithm exceeds 7200 seconds, we interrupt it and record the runtime as 7200 seconds.

\paragraph{Performance with different $n$.} We let $n = \{5, 10, 30,$ $50, 100, 200, 300, 400, 500, 600, 700, 800, 900, 1000\}$, and let $\rp$ and $\rv$ be $\lfloor \frac{n}{2}\rfloor$. As shown in \cref{subfigure:n-runtime}, the runtime of TDBS is significantly lower than the other two algorithms. Although it is not an exact algorithm, the results with precision $10^{-3}$ are accurate enough for practical applications with the guarantee proven in \cref{theorem:binary_bound}. 

\paragraph{Performance with different $\rp,\rv$.} We then fix $n=100$ and let $\rp = \rv$ be $\{0,1,\dots,50\}$. As shown in \cref{subfigure:rv-runtime}, both of our proposed algorithms' runtimes are stable with different $\rp$ and $\rv$. However, the performance of MILP is highly unstable due to its lack of time complexity guarantee.

Additional experiments are presented in \cref{app:additional_experiments}.

\section{Case Study on Anti-poaching}
\label{sec:case_study}

\begin{figure*}[t]
    \centering
    
    \begin{subfigure}{0.12\textwidth}
      \centering
      \includegraphics[width=\linewidth]{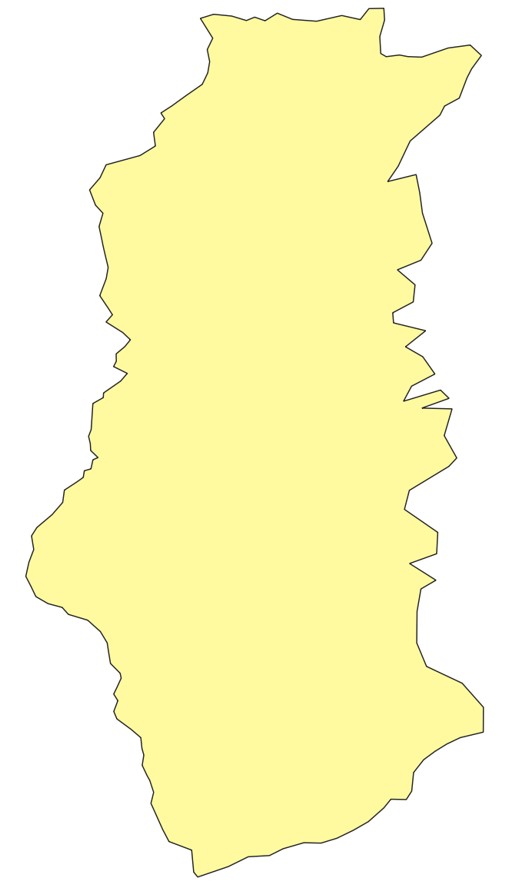}
      \caption{Contour}
      \label{subfigure:contour}
    \end{subfigure}
    \hfill
    \begin{subfigure}{0.285\textwidth}
      \centering
      \includegraphics[width=\linewidth]{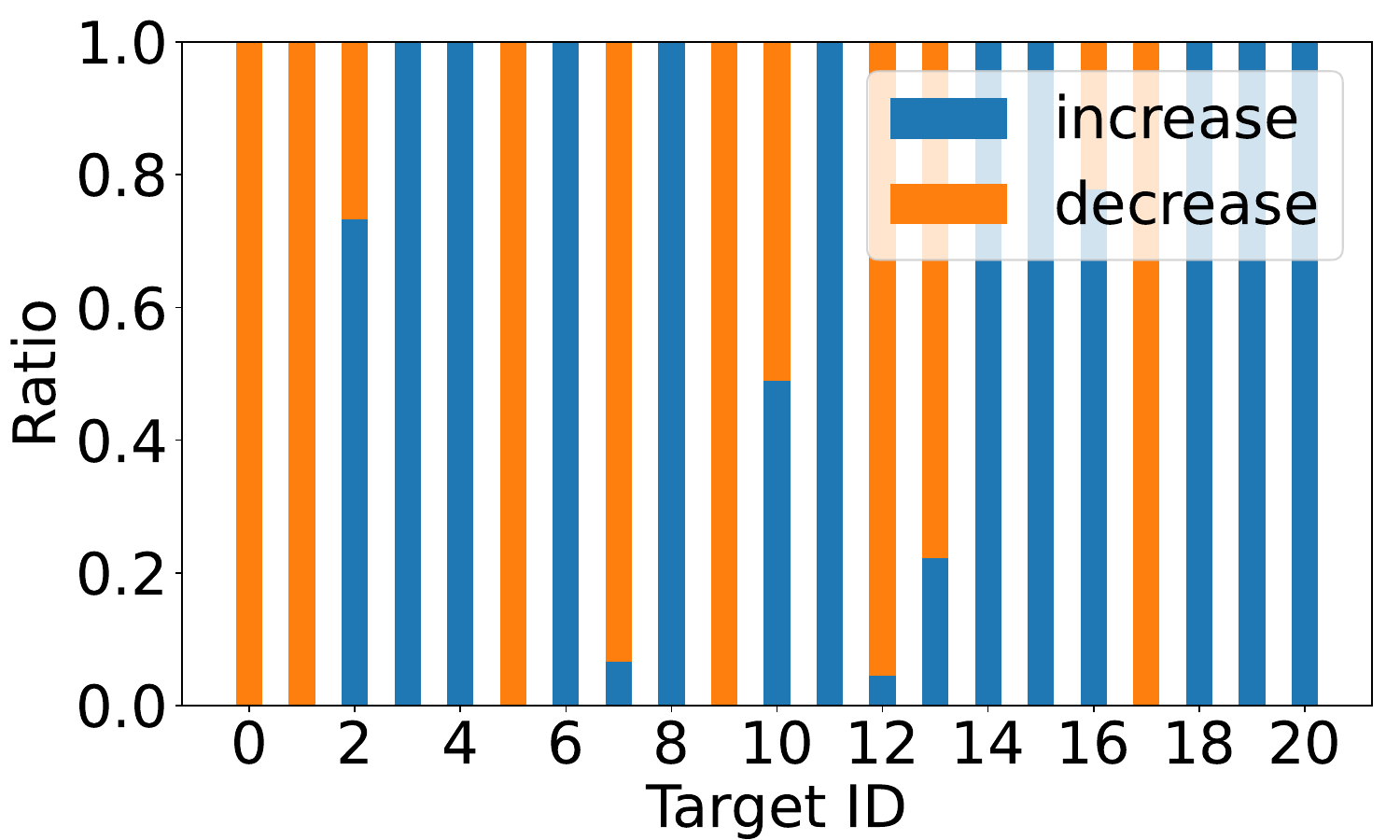}
      \caption{Ratio of increase and decrease}
      \label{subfigure:ratio}
    \end{subfigure}
    \hfill
    \begin{subfigure}{0.285\textwidth}
      \centering
      \includegraphics[width=\linewidth]{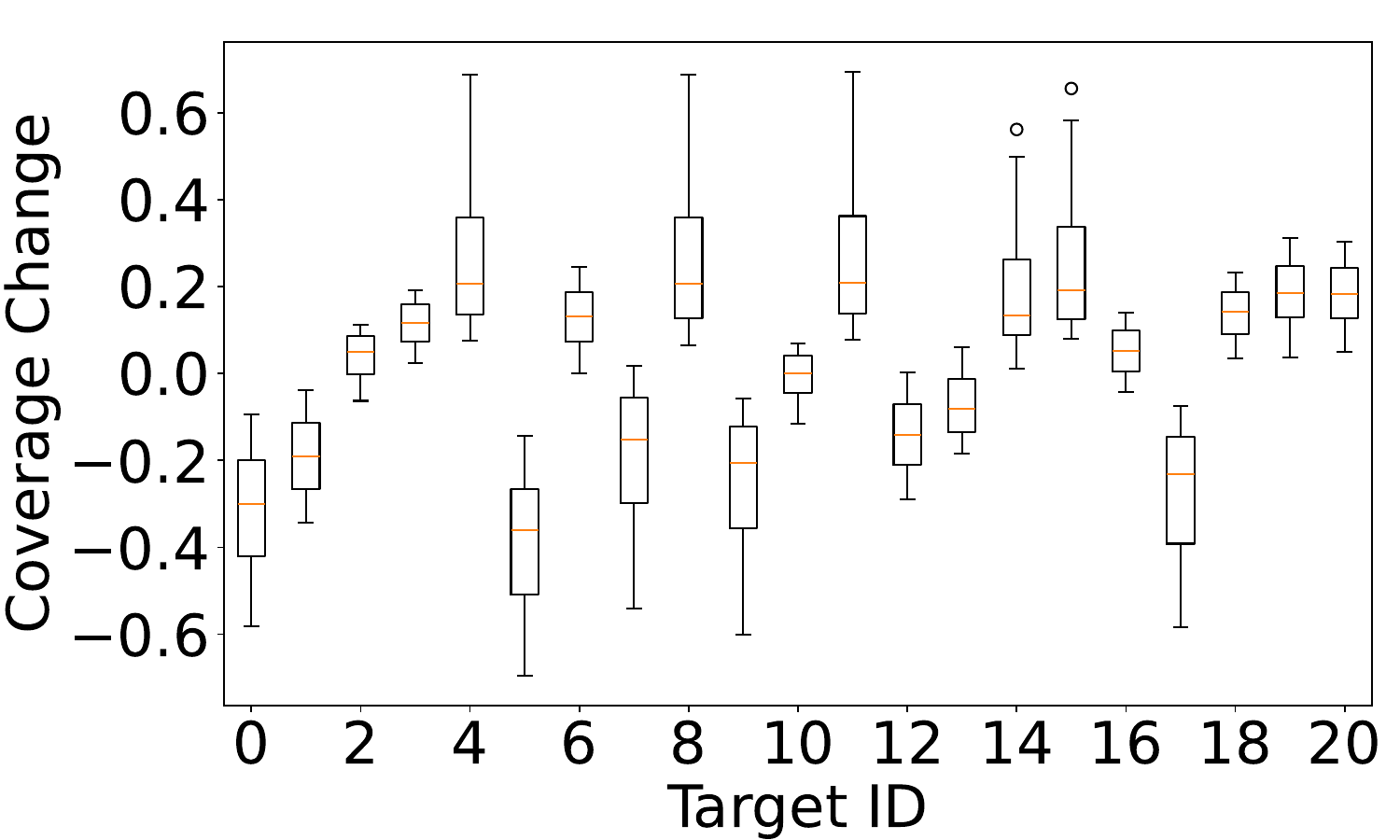}
      \caption{Coverage change}
      \label{subfigure:coverage_change}
    \end{subfigure}
    \begin{subfigure}{0.285\textwidth}
        \centering
        \includegraphics[width=\linewidth]{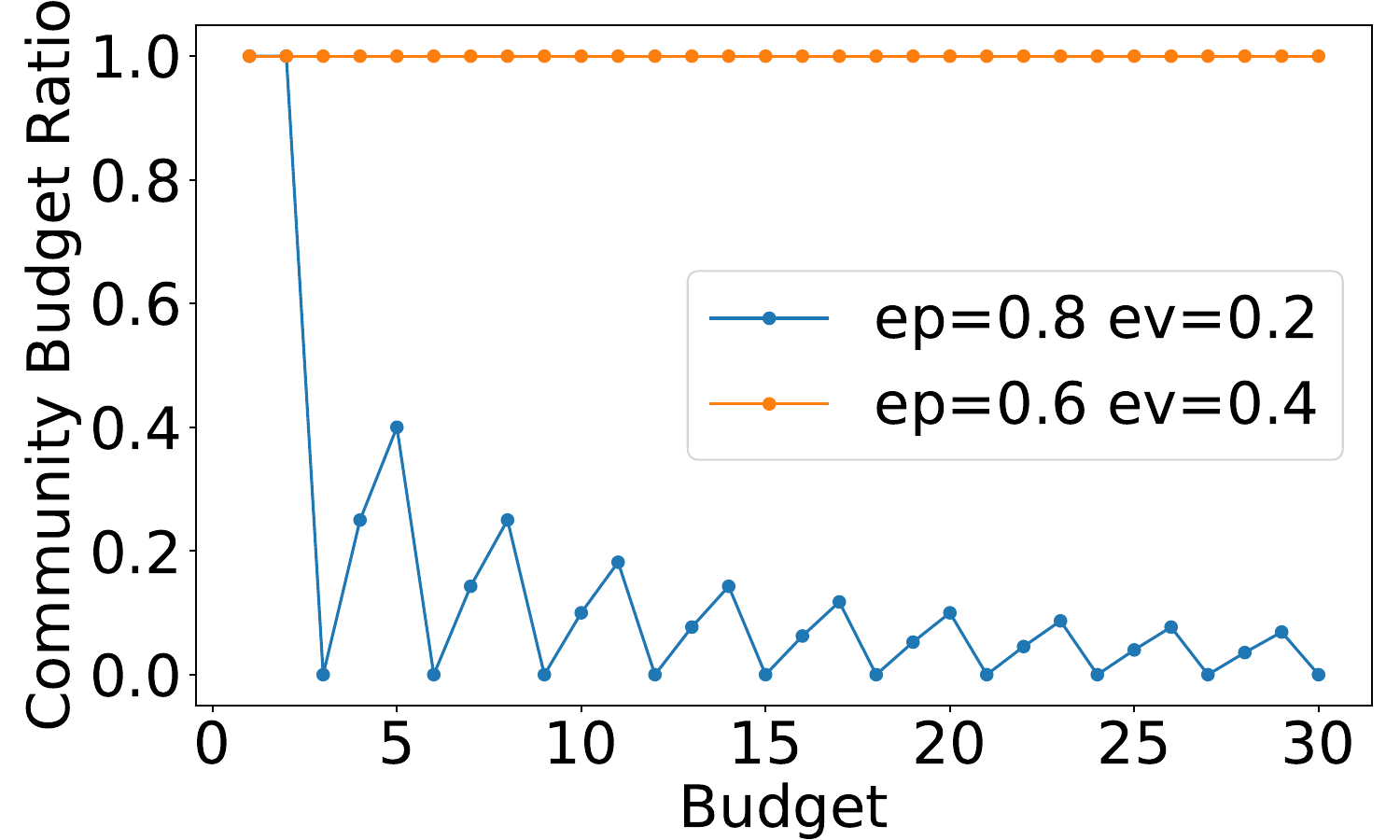}
        \caption{Budget allocation}
        \label{subfigure:budget}
    \end{subfigure}

    \caption{The contour of the studied forest farm and case study results regarding advice on strategies and budget allocation.}
    \label{fig:case_study}
\end{figure*}

We have applied RACPP to a protected area in Northeast China, home to the Manchurian tiger. To protect the raw data which includes past patrol allocation and animal density, we only report the information that can be made public based on discussions with local agencies. The contour of the protected area is shown in \cref{subfigure:contour}. The protected area is divided into 21 2km$\times$2km regions in the last three patrol seasons for patrol planning, and we naturally use these regions as targets, i.e., $T = \{0, 1, \dots, 20\}$. Ranger resources $\rp$ and villager resources $\rv$ are provided by local agencies. Besides, since roe deer and sika deer are the main prey of Manchurian tigers in that area, we use a weighted distribution of multiple species including roe deer, sika deer, and wild boars to estimate the reward for the poacher $\ra_{i}$, which is shown in \cref{table:target_value} in \cref{app:additional_case_study}. We set $\pd_{i}=-\ra_{i}$. Since defenders care less about where they find snares or catch attackers and attackers face the same amount of fine wherever they are caught, $-\pa_{i}$ and $\rd_{i}$ are set to a fixed number 10.

In this experiment, we calculate and compare the optimal patrol strategy with the current one. First, we obtain the villager allocation strategy from local agencies and process last season's ranger patrol records to determine their strategy. Specifically, we calculate the total length of patrol routes within each target to derive the distribution of ranger efforts and, combined with ranger resources $\rp$, estimate the resources allocated to each target.
Since the exact values of $\ep$ and $\ev$ are unknown, we enumerate their values in ${0.1, 0.2, \dots, 0.9}$ with $\ep \geq \ev$, resulting in 45 settings. After generating optimal defender strategies and total coverage for each target, we compare them to the current strategy in the protected area and summarize the findings in \cref{subfigure:ratio}. The length of the orange bar shows how many $(\ep, \ev)$ settings suggest reducing coverage on that target. For instance, for target 0, the optimal coverage is lower than the current level across all settings, while target 18 consistently requires increased coverage. \cref{subfigure:coverage_change} shows the distribution of coverage changes for all targets. Defender utility is expected to improve 25. 9\% - 152. 6\%, with an average of 83. 1\%.

When the total budget for recruiting rangers and villagers increases, should we use the extra money to recruit more rangers or more villagers? The current cost ratio of one ranger versus one villager is 3:1. 
We calculate the optimal plan of allocating the extra money with two different $(\ep, \ev)$ settings when the budget increase ranges from 1 unit to 30 units and show the results in \cref{subfigure:budget}. When $\ep = 0.8$ and $\ev = 0.2$, prioritizing recruitment of rangers is more cost-efficient. In contrast, when $\ep = 0.6$ and $\ev = 0.4$, we should spend all budgets on villagers. Additional results for other settings are shown in \cref{app:additional_case_study}.

We run additional experiments that take into account the terrain information as it is relatively easier to find snares on targets with higher slope variance according to domain experts. The results are shown in \cref{app:additional_case_study}. 

We plan to collaborate with the local forest bureau in Northeast China to adjust future patrol resource allocation based on the case study findings. The planned work includes refining the range of $\ep$ and $\ev$ through a poaching detection competition where the locations of the poaching tools are known and thus offering more precise recommendations.

\newpage

\bibliography{ref}

\newpage
\appendix

\section{Missing Proofs in \cref{subsec:two_dimensional_binary_search}}
\label{app:missing_proofs_tdbs}

\subsection{Proof of \cref{lemma:binary_judge_correctness}}
\label{appsub:proof_of_lemma_binary_judge}

\BinaryJudgeCorrectness*

\begin{proofof}{\cref{lemma:binary_judge_correctness}}
For a valid defender strategy profile $(\b p, \b v)$ satisfying $p_{i^{*}} = p, v_{i^{*}} = v$ and $i^{*}$ is the attacker's best response, the followings must hold: (a) $p_{i}\in \mathbb R_{\geq 0}$, (b) $v_{i} \in \mathbb{N}$, (c) $\sum_{i \in T}p_i\leq \rp$, (d) $\sum_{i \in T}v_i\leq \rv$, (e) $p_{i^{*}} = p$, (f) $v_{i^{*}} = v$, and (g) $\ua_{i^{*}} \geq \ua_{i}$ $(\forall i \in T)$.

    \textbf{Sufficiency.} After \cref{alg:binary_judge}, we construct $(\b p, \b v)$ as 
    \begin{align*}
        p_{i} &= \left\{\begin{array}{ll}
            \delta_{i} / \ep & (i \in T \setminus \{i^{*}\}) \\
            p & (i = i^{*}),
        \end{array}\right. \\
        v_{i} &= \left\{\begin{array}{ll}
            v_{\mathrm{cnt}, i} + 1 & (i \in T \setminus \{i^{*}\}, i \text{ is visited on Line 12}) \\
            v_{\mathrm{cnt}, i} & (i \in T \setminus \{i^{*}\}, i \text{ is not visited on Line 12}) \\
            v & (i = i^{*}).
        \end{array}\right.
    \end{align*}
    
    We first prove that if \cref{alg:binary_judge} returns True, the constructed $(\b p, \b v)$ is a valid defender strategy profile that satisfies (a) to (g). 

    After executing Lines 6 to 10, $\delta_{i} \geq 0$ since $v_{\mathrm{cnt}, i} \leq c_{\mathrm{min}} / \ev$. The property still holds after executing Lines 11 to 12. Hence, $p_{i} = \delta_{i}/\ep \geq 0$ $(\forall i \in T \setminus \{i^{*}\})$. Combined with $p_{i^{*}} = p \geq 0$, (a) holds.

    Since $c_{\mathrm{min}} \geq 0$ and $\rv_{\mathrm{remain}} \geq 0$ always hold on Lines 6 to 10,  $v_{\mathrm{cnt}, i} = \min \left( \lfloor c_{\mathrm{min}}/\ev\rfloor, \rv_{\mathrm{remain}} \right) \in \mathbb{N}$. After executing Lines 11 to 12, the extra 1 added to some of the $v_{i}$ doesn't break the property. Combined with $v_{i^{*}} = v \in \mathbb{N}$, (b) holds.

    From Line 13, we know that $\sum_{i \in T\setminus \{i^{*}\}} \delta_i / \ep \leq \rp_{\mathrm{remain}}$. Since $\rp_{\mathrm{remain}} = \rp - p_{i^{*}}$, we derive that $\sum_{i \in T }p_i \leq \rp$ and (c) holds.

    From Lines 8 to 9, $\rv_{\mathrm{remain}} + \sum_{i \in T\setminus \{i^{*}\}} v_{\mathrm{cnt}, i }= \rv - v_{i^{*}}$ always holds. After executing Lines 11 to 12, the sum of extra 1 added to some of the $v_{i}$ doesn't exceed $\rv_{\mathrm{remain}}$. Therefore, $\sum_{i \in T}v_i - v_{i^{*}} \leq \rv - v_{i^{*}}$, and (d) holds.

    (e) and (f) hold since we set $p_{i^{*}}$ and $v_{i^{*}}$ to $p$ and $v$.

    On Line 7, we calculate a minimum coverage $c_{i}$ as $c_{\mathrm{min}, i}$ to ensure $\ua_{i} \leq \ua_{i^{*}}$. After executing Lines 6 to 10, $\delta_{i} + v_{\mathrm{cnt}, i} \cdot \ev = c_{\mathrm{min}, i}$ and $\delta_{i} < \ev$ if $\rv_{\mathrm{remain}} > 0$. After executing Lines 11 to 12,  $\delta_{i} + v_{i} \cdot \ev \geq c_{\mathrm{min}, i}$ since some of the $\delta_{i}$ are replaced by extra $\ev$.  Hence, $p_{i} \cdot \ep + v_{i} \cdot \ev \geq c_{\mathrm{min}, i}$ $(\forall i \in T \setminus \{i^{*}\})$ and (g) holds.
    
    Therefore, there exists a valid defender strategy profile $(\mathbf p, \mathbf v)$ such that $p_{i^{*}} = p, v_{i^{*}} = v$ and $i^{*}$ is the attacker's best response if \cref{alg:binary_judge} returns True.

    \textbf{Necessity.} We will prove the necessity of \cref{alg:binary_judge} by contradiction. Assume that there exists a valid defender strategy profile $(\b p,\b v)$ such that $p_{i^{*}} = p, v_{i^{*}} = v$ and $i^{*}$ is the attacker's best response, while \cref{alg:binary_judge} returns False.
    
    Although \cref{alg:binary_judge} returns False, it generates a defender strategy $\b v'$ for villagers, where
    $$
    v'_{i} = \left\{\begin{array}{ll}
        v_{\mathrm{cnt}, i} + 1 & (i \in T \setminus \{i^{*}\}, i \text{ is visited on Line 12}) \\
        v_{\mathrm{cnt}, i} & (i \in T \setminus \{i^{*}\}, i \text{ is not visited on Line 12}) \\
        v & (i = i^{*}).
    \end{array}\right.
    $$ 
    The strategy $\b v'$ satisfies properties (b), (d) and (f) with an argument similar to the sufficiency part.
    
    Without loss of generality, we can assume both $\b v$ and $\b v'$ allocate all villagers to the targets. Hence, we know that $\sum_{i \neq i^{*}} v_{i} = \sum_{i \neq i^{*}} v'_{i}$.
    
    After executing Lines 6 to 10, \cref{alg:binary_judge} allocates $v_{\mathrm{cnt}, i}$ villagers, which don't exceed $c_{\mathrm{min}, i} / \ev$, thus $\cw_{i}(\b v') = 0$ $(\forall i \in T)$. On Lines 11 to 12, \cref{alg:binary_judge} distributes villages to targets with maximum $\delta_{i} = c_{\mathrm{min}, i} - v_{\mathrm{cnt}, i} \cdot \ev$, minimizing $v_{i} \cdot \ev - c_{\mathrm{min}, i}$. Hence, $\b v'$ guarantees a minimum $\sum_{i \neq i^{*}}\cw_{i}(\b v')$ and we can know that $\sum_{i \neq i^{*}}\cw_{i}(\b v') \leq \sum_{i \neq i^{*}}\cw_{i}(\b v)$. 
    
    Therefore, after villager distributions are set, the sum of coverage needed to by filled with rangers over all targets $\sum_{i \neq i^{*}}(c_{\mathrm{min}, i} - v'_{i} \cdot \ev + \cw_{i}(\b v')) \leq \sum_{i \neq i^{*}}(c_{\mathrm{min}, i} - v_{i} \cdot \ev + \cw_{i}(\b v))$. Since \cref{alg:binary_judge} returns False, we derive that $\sum_{i \neq i^{*}}(c_{\mathrm{min}, i} - v'_{i} \cdot \ev + \cw_{i}(\b v')) > (\rp - p) \cdot \ep$. Therefore, we can conclude that $(\rp - p) \cdot \ep < \sum_{i \neq i^{*}}(c_{\mathrm{min}, i} - v_{i} \cdot \ev + \cw_{i}(\b v))$, which means there are no enough ranger efforts to construct a $\b p$ as the strategy for ranger in the valid defender strategy profile $(\b p, \b v)$. We get a contradiction.
\end{proofof}

\subsection{Proof of \cref{lemma:mono_patroller_and_villager}}
\label{appsub:proof_of_lemma_mono_patroller_and_villager}

\MonoPatrollerAndVillager*

\begin{proofof}{\cref{lemma:mono_patroller_and_villager}}
    We can construct a defender strategy profile $(\b p', \b v')$  where
    \begin{align*}
        p'_{i} &= \left\{\begin{array}{ll}
            p_{i} & (i \in T \setminus \{i^{*}\}) \\
            p & (i = i^{*}),
        \end{array}\right. \\
        v'_{i} &= \left\{\begin{array}{ll}
            v_{i} & (i \in T \setminus \{i^{*}\}) \\
            v & (i = i^{*}).
        \end{array}\right.
    \end{align*}
    We will show that $(\b p', \b v')$ satisfies (a) $p'_{i}\in \mathbb R_{\geq 0}$, (b) $v'_{i} \in \mathbb{N}$, (c) $\sum_{i \in T}p'_i\leq \rp$, (d) $\sum_{i \in T}v'_i\leq \rv$, (e) $p'_{i^{*}} = p$, (f) $v'_{i^{*}} = v$, and (g) $\uap_{i^{*}} \geq \uap_{i}$ $(\forall i \in T)$.

    Since $(\b p, \b v)$ is a valid defender strategy profile, we know that $p_{i}\in \mathbb R_{\geq 0}$, $v_{i} \in \mathbb{N}$, $\sum_{i \in T}p_i\leq \rp$ and $\sum_{i \in T}v_i\leq \rv$. Combined with $0 \leq p'_{i^{*}} = p \leq p_{i^{*}}$ and $0 \leq v'_{i^{*}} = v \leq v_{i^{*}}$ $(v \in \mathbb{N})$, we can easily derive that (a), (b), (c) and (d) hold. (e) and (f) hold because we set $p'_{i^{*}}$ and $v'_{i^{*}}$ respectively to $p$ and $v$. 
    
    We know that $\ua_{i^{*}} \geq \ua_{i}$ $(\forall i \in T)$ since with the defender strategy profile $(\b p, \b v)$, an attacker will attack target $i^{*}$. Because $p'_{i} = p_{i}$ $(\forall i \in T \setminus \{i^{*}\})$ and $v'_{i} = v_{i}$ $(\forall i \in T \setminus \{i^{*}\})$, we can get that $\uap_{i} = \ua_{i}$ $(\forall i \in T \setminus \{i^{*}\})$. We can derive that $\uap_{i^{*}} \geq \ua_{i^{*}}$ from the fact that $p'_{i^{*}} \leq p_{i^{*}}$ and $v'_{i^{*}} \leq v_{i^{*}}$. We can conclude that $\uap_{i^{*}} \geq \ua_{i^{*}} \geq \ua_{i} = \uap_{i}$ $(\forall i \in T \setminus \{i^{*}\})$. Hence, property (g) also holds.
\end{proofof}

\subsection{Proof of \cref{lemma:mono_coverage_unchanged}}
\label{appsub:proof_of_lemma_mono_coverage_unchanged}

\MonoCoverageUnchanged*

\begin{proofof}{\cref{lemma:mono_coverage_unchanged}}
    We can construct a defender strategy profile $(\b p', \b v')$  where $p'_{i^{*}} = p$ and $v'_{i^{*}} = v$, so $(v'_{i^{*}} - v_{i^{*}}) \cdot \ev = (p_{i^{*}} - p'_{i^{*}}) \cdot \ep$. For $i \in T \setminus \{i^{*}\}$, we can keep $c_{i} = c'_{i}$ by replacing the $(v'_{i^{*}} - v_{i^{*}})$ villagers with the spared $(p_{i^{*}} - p'_{i^{*}})$ ranger efforts. Therefore, we can derive that $\uap_{i} = \ua_{i}$ $(\forall i \in T)$. The adjustment from $(\b p,\b v)$ to $(\b p', \b v')$ doesn't break the definition of a valid defender strategy profile.
    
    Since with the defender strategy profile $(\b p, \b v)$, target $i^{*}$ is the attacker's best response, we know that $\ua_{i^{*}} \geq \ua_{i}$  $(\forall i \in T)$. We can conclude that $\uap_{i^{*}} = \ua_{i^{*}} \geq \ua_{i} = \uap_{i}$ for all $i \in T \setminus \{i^{*}\}$. Hence, target $i^{*}$ is still the attacker's best response with valid defender strategy profile $(\b p',\b v')$.
\end{proofof}

\subsection{Proof of \cref{theorem:binary_bound}}
\label{appsub:proof_of_theorem_binary_bound}

\BinaryBound*

\begin{proofof}{\cref{theorem:binary_bound}}
First, we will prove the correctness of \cref{alg:binary}. Let target $i^{*}$ be the attacker's best response with the defender strategy profile $(\b p, \b v)$, and $i^{*\prime}$ be that with $(\b p^*, \b v^*)$. We can prove the correctness by discussing the two cases whether $i^{*}$ is equal to $i^{*\prime}$ or not.

    In the first case, we discuss the situation where $i^{*} = i^{*\prime}$. By applying \cref{lemma:mono_coverage_unchanged} to $\b v^{*}$, replacing every $\ev / \ep$ ranger efforts on target $i^{*\prime}$ with one villager won't cause utility to decrease. Therefore, we can assume that $v_{i^{*}} = v_{i^{*\prime}}$ after executing Lines 5 to 12 and finding the maximum $v_{i^{*}}$ that can be allocated to target $i^{*}$. By \cref{lemma:mono_patroller_and_villager},  given target $i^{*}$ and villager resources $v_{i^{*}}$, if $i^{*}$ is not the attacker's best response when $x$ ranger efforts are allocated on target $i^{*}$, we can reject all ranger efforts $p$ larger than $x$. Therefore, binary search on Lines 13 to 20 can gradually narrow the range that $p_{i^{*}}$ can fall in given $i^{*}$ and $v_{i^{*}}$. Combined with \cref{lemma:binary_judge_correctness}, we can know that $p^{*}_{i^{*\prime}} - p_{i^{*}} < \varepsilon$. On account of $v_{i^{*}} = v_{i^{*\prime}}^{*}$, coverage $c^{*}_{i^{*\prime}} - c_{i^{*}} < \ep \cdot \varepsilon$. Because of the monotonicity of defender utility on coverage, we can conclude that $u(\b p^{*}, \b v^{*}) - u(\b p, \b v) < \ep \cdot \varepsilon \cdot (\rd_{i^{*\prime}} - \pd_{i^{*\prime}}) \leq \ep \cdot 2M\varepsilon$.

    If $i^{*} \neq i^{*\prime}$, considering \cref{alg:binary} chooses a strategy with maximum utility on Line 22, we know that $u(\b p, \b v)$ is larger than any defender utility that can be achieved when $i^{*} = i^{*\prime}$. Integrated with the fact proved in the first case, we can claim that $u(\b p^{*}, \b v^{*}) - u(\b p, \b v) < \ep \cdot 2M\varepsilon$ if $i^{*} \neq i^{*\prime}$.

    Then, we will prove the time complexity of \cref{alg:binary} is $O(n^{2} \cdot \log\frac{M}{\varepsilon})$. In \cref{alg:binary}, there are totally $O(n(1 + \log \rv + \log\frac{\rp}{\varepsilon})) = O(n \cdot \log\frac{M}{\varepsilon})$ calls to \cref{alg:binary_judge}. In \cref{alg:binary_judge}, Lines 2 to 3 and Lines 6 to 10 are of time complexity $O(n)$; Lines 11 to 12 are also of time complexity $O(n)$ since we can expect to find the $\min(n, \rv_{\mathrm{remain}})$ largest elements in $\boldsymbol{\delta}$ in  linear time with quickselect and quicksort \cite{cormen2022introduction}. Therefore, \cref{alg:binary_judge} works in $O(n)$ and \cref{alg:binary} is of time complexity $O(n^{2} \cdot \log\frac{M}{\varepsilon})$.
\end{proofof}

\section{Missing Details in \cref{subsec:exact_algorithm}}
\label{app:details_exact}

\subsection{Details of the HW Algorithm}

In this section, we will present some details of the HW algorithm, which is omitted in \cref{subsec:exact_algorithm}. It will include a formal description of the HW algorithm and a rigorous proof of its correctness and time complexity.

Given the villager strategy $v_{i}$ on target $i$, the coverage caused by villagers is defined as $\cv_{i}$, and the attacker utility only taking villagers into account is $\uav_{i}$. The value of $\cv_{i}$ and $\uav_{i}$ can be computed with
    
\begin{equation}
    \begin{aligned}
         \cv_{i} &= \min(\ev \cdot v_{i}, 1)  \\
         \uav_{i} &= \ra_{i} \cdot (1 - \cv_{i}) + \pa_{i} \cdot \cv_{i}
        \label{equation:v_to_uav}
     \end{aligned}
\end{equation}

In \cref{subsec:exact_algorithm}, we have mentioned that swaps are triggered at critical points, where for targets $i,j$, the coverage resulted from rangers on target $i$ is equal to the coverage (above the sea level) caused by the last villager on target $j$. This equality relationship can be formally described as
\begin{equation}
         \frac{\uav_{i} - \uap_{i}}{\ra_{i} - \pa_{i}} = \frac{\ra_{j} - (\ra_{j} - \pa_{j} \cdot \ev \cdot (v_{j} - 1)) - \uap_{i}}{\ra_{j} - \pa_{j}}
    \label{equation:water_line_boundary_process}
\end{equation}
where $\uap_{i}$ is the sea level at the critical point. Consider the process of Waterfilling starting from a sea level $\ua_{i}$, for a specific pair of targets $i,j$, the minimum sea level drop before hitting a critical point can be calculated as
\begin{equation}
    \begin{aligned}
         &u_{\mathrm{drop}}(i, j) = \ua_{i} - \uap_{i} = \ua_{i} - \uav_{i}  \\
         &+ \frac{(\ra_{j} - (\ra_{j} - \pa_{j}) \cdot \ev \cdot (v_{j} - 1) - \uav_{i}) \cdot (\ra_{i} - \pa_{i})}{(\ra_{j} - \pa_{j}) - (\ra_{i} - \pa_{i})}
         \label{equation:water_line_boundary}
     \end{aligned}
\end{equation}

To find the minimum sea level drop before triggering a critical point, we need to calculate $u_{\mathrm{drop}}(i, j)$ for all pairs of targets $i,j$, where $i$ is in the critical set and $j$ is not in the set, and find the minimum value. The formal steps are presented in \cref{alg:swap_line}. Here, $S_{\rm{cur}}$ is the current critical set, and $i^{*}$ is the target that the attacker will attack.

\begin{algorithm}[htbp]
    \caption{Calculate min utility drop before swapping}
    \label{alg:swap_line}
    
    \textbf{Output}: Minimum utility drop, the target swapping out rangers and the target swapping out villagers

    \begin{algorithmic}[1]
         \STATE \textbf{procedure} getSwapLine()
         \STATE Let $(u_{\mr{change}}, i_{\mr{outp}}, i_{\mr{outv}}) \gets (\infty, \mr{None}, \mr{None})$.
         \FOR{$i \in S_{\mr{cur}} \setminus \{i^{*}\}$}
            \FOR{$j \in T \setminus (S_{\mr{cur}} \cup \{i^{*}\})$}
                \IF{$u_{\mr{drop}}(i, j) < u_{\mr{change}}$ \textbf{and} $u_{\mr{cur}} - u_{\mr{drop}}(i, j) \geq \pa_{j}$ \textbf{and} $w_{j} < w_{i}$ \textbf{and} $p_{j} = 0$ \textbf{and} $v_{j} > 0$}
                    \STATE Let $(u_{\mr{change}}, i_{\mr{outp}}, i_{\mr{outv}}) \gets (u_{\mr{drop}}(i, j), i, j)$.
                \ENDIF
            \ENDFOR
         \ENDFOR        
        \RETURN $u_{\mr{change}}, i_{\mr{outp}}, i_{\mr{outv}}$.
    \end{algorithmic}
\end{algorithm}

Then in \cref{alg:exact_patroller}, we can solve the important subproblem that given a target $i^*$ that the attacker will attack and a villager strategy $v_{i^*}$ on $i^*$, we need to allocate the remaining villagers and rangers to maximize the defender utility.

\begin{algorithm}
    \caption{Ranger strategy generator of HW}
    \label{alg:exact_patroller}

    \textbf{Input}: Input instance $\I$, target $i^{*}$, villager strategy $v_{i^{*}}$on $i^{*}$

    \textbf{Output}: Optimal valid strategy profile $(\b p, \b v)$

    \begin{algorithmic}[1]
        \STATE Compute $\ua_{i^{*}}$ using \eqref{equation:pv_to_ua} with $(0, v_{i^{*}})$.
        \STATE Let $\ua_{i} \gets \ra_{i}$ \textbf{for all} $i \in T, i \neq i^{*}$.
        \STATE Let $v_{i} \gets 0$ \textbf{for all} $i \in T, i \neq i^{*}$.
        \STATE Let $p_{i} \gets 0$ \textbf{for all} $i \in T$.
        \STATE Let $w_{i} \gets 1 / (\ra_{i} - \pa_{i})$ \textbf{for all} $i \in T$.
        \STATE Let $(\rp_{\mr{remain}}, \rv_{\mr{remain}}) \gets (\rp, \rv - v_{i^{*}})$.
        \FOR{$t \in \{1, 2, \dots, \rv_{\mr{remain}}\}$}
            \STATE Let $j \gets \arg\max_i \{\ua_{i}\mid i\ne i^*,\ua_{i} \neq \pa_{i}\}$.
            \STATE Let $v_{j} \gets v_{j} + 1$ and update $\ua_{j}$ using \eqref{equation:pv_to_ua} with $(0, v_{j})$.
        \ENDFOR
        \STATE Let $\uav_{i} \gets \ua_{i}$ \textbf{for all} $i \in T$.
        \WHILE{$\rp_{\mr{remain}} > 0$ \textbf{and} $\exists i, \ua_{i} \neq \pa_{i}$}
            \IF{$\exists i \textbf{ such that } \ua_{i} = \pa_{i} = \ua_{i^{*}}$}
            \RETURN $\b p$.
            \ENDIF
            \STATE Let $u_{\mr{last}} \gets u_{\mr{cur}}$ \textbf{if} $u_{\mr{cur}}$ exists.
            \STATE Let $u_{\mr{cur}} \gets \max_i\{\ua_{i} \mid \ua_{i} \neq \pa_{i} \}$.
            \STATE Let $u_{\mr{next}} \gets \max_i\{\ua_{i} \mid \ua_{i} \ne u_{\mr{cur}}, \ua_{i} \neq \pa_{i}\}$.
            \STATE Let $S_{\mr{cur}} \gets \{i \mid \ua_{i} = u_{\mr{cur}}, \ua_{i} \neq \pa_{i}\}$.
            \STATE Let $f_{\mr{swap}} \gets \mr{True}$.
            \STATE Let $(u_{\delta}, i_{\mr{outp}}, i_{\mr{outv}}) \gets$ getSwapLine().
            \STATE Let $\pa_{\mr{max}} \gets \max_i\{\pa_{i} \mid i \in T\}\textbf{ if }i^{*} \in S_{\mr{cur}}$.
            \STATE Let $\pa_{\mr{max}} \gets \max_i\{\pa_{i} \mid i \in S_{\mr{cur}}\}\textbf{ if }i^{*} \not\in S_{\mr{cur}}$.
            \IF{$u_{\mr{cur}} - u_{\delta} < \pa_{\mr{max}}$}
                \STATE Let $u_{\delta} \gets u_{\mr{cur}} - \pa_{\mr{max}}$.
                \STATE Let $f_{\mr{swap}} \gets \mr{False}$.
            \ENDIF
            \IF{$u_{\mr{next}}\ne\mr{None}$ \textbf{and} $u_{\mr{cur}} - u_{\delta} < u_{\mr{next}}$}
                \STATE Let $u_{\delta} \gets u_{\mr{cur}} - u_{\mr{next}}$.
                \STATE Let $f_{\mr{swap}} \gets \mr{False}$.
            \ENDIF
            \STATE Let $\rp_{\delta} \gets \sum_{i \in S_{\mr{cur}}}w_{i} \cdot u_{\delta}/ \ep $.
            \IF{$\rp_{\delta} > \rp_{\mr{remain}}$}
                \STATE Let $\rp_{\delta} \gets \rp_{\mr{remain}}$.
                \STATE Let $f_{\mr{swap}} \gets \mr{False}$.
            \ENDIF
            \STATE Let $u_{\delta} \gets \rp_{\delta} \cdot \ep / (\sum_{i \in S_{\mr{cur}}}w_{i})$.
            \STATE Let $\rp_{\mr{remain}} \gets \rp_{\mr{remain}} - \rp_{\delta}$.
            \FOR{$i \in S_{\mr{cur}}$}
                \STATE Let $\ua_{i} \gets \ua_{i} - u_{\delta}$.
                \STATE Let $p_{\delta} \gets u_{\delta} / (\ep \cdot (\ra_{i} - \pa_{i}))$.
                \STATE Let $p_{i} \gets p_{i} + p_{\delta}$.
            \ENDFOR
            \IF{$f_{\mr{swap}}$}
                \STATE Let $v_{i_{\mr{outv}}} \gets v_{i_{\mr{outv}}} - 1$.
                \STATE Let $v_{i_{\mr{outp}}} \gets v_{i_{\mr{outp}}} + 1$.
                \STATE Let $p_{i_{\mr{outv}}} \gets p_{i_{\mr{outp}}}$.
                \STATE Let $p_{i_{\mr{outp}}} \gets 0$.
                \FOR{$i \in \{i_{\mr{outv}}, i_{\mr{outp}}\}$}
                    \STATE Update $\ua_{i}$ and $\uav_{i}$ using \eqref{equation:pv_to_ua}, \eqref{equation:v_to_uav} with $(p_{i}, v_{i})$.
                \ENDFOR
            \ENDIF
        \ENDWHILE
        \RETURN $(\b p, \b v)$.
    \end{algorithmic}
\end{algorithm}

Finally, we state the HW algorithm in \cref{alg:exact}.

\begin{algorithm}[htbp]
    \caption{Hybrid Waterfilling Algorithm}
    \label{alg:exact}

    \textbf{Input}: Input instance $\I$

    \textbf{Output}: Maximum defender utility

    \begin{algorithmic}[1]
         \STATE Let $u_{\mathrm{ans}} \gets -\infty$.
        \FOR{$i^{*} \in \{0, 1, \dots, n-1\}$}
            \IF{\cref{alg:binary_judge} returns $\mathrm{False}$ on $(\I, i^{*}, 0, 0)$}
                \STATE \textbf{continue}
            \ENDIF
            \STATE Let $(v_{\mathrm{left}},v_{\mathrm{right}}) \gets (0,\rv)$.
            \WHILE{$v_{\mathrm{left}} \leq v_{\mathrm{right}}$}
                \STATE Let $v_{\mathrm{cur}} \gets \lfloor(v_{\mathrm{left}} + v_{\mathrm{right}})/{2}\rfloor$.
                 \IF{\cref{alg:binary_judge} returns $\mathrm{True}$ on $(\I, i^{*}, 0, v_{\mathrm{cur}})$}
                    \STATE Let $v_{\mathrm{left}} \gets v_{\mathrm{cur}} + 1$.
                    \STATE Let $v_{i^{*}} \gets v_{\mathrm{cur}}$.
                \ELSE
                    \STATE Let $v_{\mathrm{right}} \gets v_{\mathrm{cur}} - 1$.
                \ENDIF
            \ENDWHILE
            \STATE Let $(\b p, \b v) \gets$ result of \cref{alg:exact_patroller} on $(\I, i^{*}, v_{i^{*}})$.
            \STATE Compute $\ud_{i^{*}}$ using $(p_{i^{*}}, v_{i^{*}})$.
            \STATE Let $u_{\mathrm{ans}} \gets \max(u_{\mathrm{ans}}, \ud_{i^{*}})$.            
        \ENDFOR
        \RETURN $u_{\mathrm{ans}}$.
    \end{algorithmic}
\end{algorithm}

\subsection{Proofs of the HW Algorithm}

Then, we will demonstrate the correctness and time complexity of the HW algorithm. After all villagers are allocated and during the process of Waterfilling, there is an important property that on every target, there is at most one wasted villager. Formally, we have the following lemma.

\begin{lemma}
\label{lemma:villager_wasted_bound}
    Let $u_{\mr{cur}}$ be $\max_i\{\ua_{i} \mid \ua_{i} \neq \pa_{i} \}$. Each time \cref{alg:exact_patroller} reaches Line 11, for any target $i \neq i^{*}$, if $\ua_{i} < u_{\mr{cur}}$, then $\ra_{i} - (\ra_{i} - \pa_{i}) \cdot \ev \cdot (v_{i} - 1) \geq u_{\mr{cur}}$.
\end{lemma}

\begin{proofof}{\cref{lemma:villager_wasted_bound}}
    First, we will prove by contradiction that when \cref{alg:exact_patroller} reaches Line 11 at the first time, every target can have at most one wasted villager. Assume that there exists a target $i$ such that $\ua_{i} < u_{\mr{cur}}$ and $\ra_{i} - (\ra_{i} - \pa_{i}) \cdot \ev \cdot (v_{i} - 1) < u_{\mr{cur}}$.

    Since $\ua_{j}$ can only decrease during Lines 7 to 9, we can know that before allocating the extra villager to $i$, sea level $u'_{\mr{cur}} \geq u_{\mr{cur}}$. That means there exists a target $k$ such that $\ua_{k} = u'_{\mr{cur}}$ and $\ua_{k} \neq \pa_{k}$. Therefore, when allocating that villager, we can derive $\ua_{k} \geq u_{\mr{cur}} > \ra_{i} - (\ra_{i} - \pa_{i}) \cdot \ev \cdot (v_{i} - 1)$. Combined with the fact that we choose $j$ with the largest $\ua_{j}$ $(\ua_{j} \neq \pa_{j})$ on Line 8, \cref{alg:exact_patroller} will allocate that villager to target $k$ instead of target $i$. There we get a contradiction.

    Then, we will prove after executing Lines 11 to 43, the property keeps. Since we move a villager when hitting a critical point, where the coverage of that villager above $u_{\mr{cur}}$ is the same as the ranger coverage on target $i_{\mr{outp}}$, the wasted villager coverage of that villager won't change if we move it to target $i_{\mr{outp}}$. Therefore, the wasted villager on every target will not be more than one villager after executing Lines 11 to 43 because we have this property before reaching Line 11.

    Hence, each time \cref{alg:exact_patroller} reaches Line 11, the property is satisfied.    
\end{proofof}

With \cref{lemma:villager_wasted_bound} in mind, we can move on to prove that the valid strategy profile $(\b p, \b v)$ generated by \cref{alg:exact_patroller} results in maximized defender utility by demonstrating it minimizes wasted villager coverage. First, \cref{lemma:local_optimal} formally describes an intuition that we can not move at most one villager and reduce the wasted villager coverage.

\begin{lemma}
    \label{lemma:local_optimal}
    
    Let $i^{*}$ be a given target. In every iteration on Lines 11 to 43 in \cref{alg:exact_patroller}, for $u \in (u_{\mr{cur}}, u_{\mr{last}}]$, \cref{alg:exact_patroller} generates a villager strategy $\b v$ such that there does not exist a $(\b v', i_{1}, i_{2})$ tuple, where $i_{1} \neq i^{*}, i_{2} \neq i^{*},$
    \begin{equation*}
        v'_{i}=\left\{\begin{array}{ll}
            v_{i} + 1 & (i = i_{1}) \\
            v_{i} - 1 & (i = i_{2}) \\
            v_{i} & (i \notin \{i_{1}, i_{2}\}),
        \end{array}\right.
    \end{equation*}    
     and $\mr{scw}(\b v', u) < \mr{scw}(\b v, u)$.
\end{lemma}

\begin{proofof}{\cref{lemma:local_optimal}}
We will explain why it is not possible to move one villager and reduce $\mr{scw}(\b v, u)$ with induction.

When reaching Line 11 the first time, there are no allocated ranger efforts. For those targets in critical set $i \in S_{\mr{cur}}$, moving a villager to $i$ will cause wasted villager coverage $\cw_{i} = \ev$. Based on \cref{lemma:villager_wasted_bound}, for targets not in critical set $j \notin S_{\mr{cur}}$, there is at most $\cw_{j} = \ev$ amount of wasted villager coverage. If $w_{i} \leq w_{j}$, the decreasing speed of $\cw_{i}$ is less than or equal to that of $\cw_{j}$, and $\cw_{i} > \cw_{j}$ will keep. If $w_{i} > w_{j}$, the decreasing speed of $\cw_{i}$ is more than that of $\cw_{j}$, and before reaching $\cw_{i} < \cw_{j}$, there is a point when $\cw_{i} = \cw_{j}$ because the change of $\cw_{i}$ and $\cw_{j}$ is continuous. At that point, \cref{alg:swap_line} hits a critical point and thereby moves a villager to a target with less wasted villager coverage.

When it is not the first time that we reach Line 11, possibly some ranger efforts have been allocated. Assume that we can not move one villager and make $\mr{scw}(\b v, u)$ smaller now, which means for any tuple $(i, j)$ such that $i \in S_{\mr{cur}}$ and $j \notin S_{\mr{cur}}$, moving a villager to $i$ and moving out all rangers on $i$ will cause wasted villager coverage $\cw_{i} > \cw_{j}$, which is the wasted villager coverage on target $j$. Similar to the proof in the last paragraph, \cref{alg:swap_line} will hit a critical point and swap before any possible situation where $\cw_{i} < \cw_{j}$ occurs.

Therefore, combine the last two paragraphs, on the basis of $\b v$, we can't move one villager and make $\mr{scw}(\b v, u)$ smaller since all such swaps have been hit before.
\end{proofof}

Based on \cref{lemma:local_optimal}, we can further present that the generated $(\b p, \b v)$ leads to the minimal wasted villager coverage.

\begin{lemma}
    \label{lemma:global_optimal}
    
    Let $i^{*}$ be a given target. In every iteration on Lines 11 to 43 in \cref{alg:exact_patroller}, for $u \in (u_{\mr{cur}}, u_{\mr{last}}]$, \cref{alg:exact_patroller} generates a villager strategy $\b v$ with a minimum $\mr{scw}(\b v, u)$.
\end{lemma}

\begin{proofof}{\cref{lemma:global_optimal}}
First, we convert the problem to another setting. Given $u$, every time we allocate one villager to a target, it will cause wasted villager coverage within $[0, \ev]$. So, every villager strategy $v$ corresponds to a set $S(\b v)$, the element of which is every villager's wasted coverage, and $| S(\b v) | = \rv$. The set of all villager position choices can be converted into a set $S_{\mr{all}}$.

From \cref{lemma:local_optimal}, we know that we cannot move one villager on the basis of $\b v$ and make $\mr{scw}(\b v, u)$ smaller. The moving out from one target is equivalent to delete the corresponding element from $S(\b v)$. Actually, we are not allowed to add a random element $m \in S_{\mr{all}} \setminus S(\b v)$ to $S(\b v)$ if the wasted villager coverage caused by $m$ is not zero and the positions with no waste on the same target have not been chosen. However, choosing position $m$ with waste is a even worse choice than choosing the position with no waste on the same target, which is allowed on account of \cref{lemma:local_optimal}. Therefore, we can derive that it is not possible to delete one element from $S(\b v)$, then add one element among not chosen positions, and reduce $\sum_{m \in S(\b v)} m$.

Then, we will prove the lemma with contradiction. Assume that there is a villager strategy $\b v'$ such that $\mr{scw}(\b v', u) < \mr{scw}(\b v, u)$, which means $\sum_{m \in S(v')} m < \sum_{m \in S(v)} m $. Since $| S(\b v') | = | S(\b v) | = \rv$, there exists $p \in S(\b v) \setminus S(\b v')$ and $q \in S(\b v') \setminus S(\b v)$ such that $q < p$. Therefore, deleting $p$ from $S(\b v)$ and then adding $q$ to it can make it smaller, which contradicts the conclusion in the last paragraph. As a result, there are no villager strategy $\b v'$ such that $\mr{scw}(\b v', u) < \mr{scw}(\b v, u)$, and the villager strategy $\b v$ achieves the minimum $\mr{scw}(\b v, u)$.
\end{proofof}

Now that we have proved the correctness of \cref{alg:exact_patroller}, which solves the important subproblem, we are ready to demonstrate the correctness and time complexity of the HW algorithm.

\ExactCorrectness*

\begin{proofof}{\cref{theorem:exact_correctness}}
    First, we will prove the correctness of \cref{alg:exact} by contradiction. Assume that there is a valid defender strategy profile $(\b p^{*}, \b v^{*})$ such that $u(\b p^{*}, \b v^{*}) > u(\b p, \b v)$. Let target $i^{*}$ be the attacker's best response with the defender strategy profile $(\b p, \b v)$, and $i^{*\prime}$ be that with $(\b p^*, \b v^*)$. Since we enumerate all $i^{*} \in \{0, 1, \dots, n-1\}$, there is a case that $i^{*} = i^{*\prime}$. According to \cref{lemma:mono_coverage_unchanged}, replacing every $\ev / \ep$ ranger efforts on target $i^{*}$ with one villager will not cause utility to decrease. Therefore, we can assume that $v^{*}_{i^{*}} = v_{i^{*\prime}}$ after executing Lines 5 to 12 and finding the maximum $v_{i^{*}}$ that can be allocated to target $i^{*}$.
    
    Based on \cref{lemma:global_optimal}, when reaching Line 45, \cref{alg:exact_patroller} generates a villager strategy $v$ with a minimum $\mr{scw}(\b v, u)$, which means we have exerted the maximum effect of villagers. So, we can assume that $\b v = \b v^{*}$. After executing \cref{alg:exact_patroller}, for any target $i \in \{i \mid p_{i} > 0, \ua_{i} \neq \pa_{i}\}$, we know that $\ua_{i} = u_{\mr{cur}}$. If $p_{i^{*}} = 0$, combined with $v_{i^{*}}$ is optimal, we can derive that $u(\b p, \b v)$ is optimal. If $\ua_{i^{*}} = \pa_{i^{*}}$, we know that $\ua_{i^{*}}$ reaches its minimum and thus $\ud_{i^{*}}$ reaches its maximum. If $\ua_{i^{*}} = u_{\mr{cur}}$, since $u(\b p^{*}, \b v^{*}) > u(\b p, \b v)$ and $i^{*} = i^{*\prime}$, we know that $p^{*}_{i^{*}} > p_{i^{*}}$. However, if we move any ranger efforts from target $i \neq i^{*}$ to target $i^{*}$, $\ua_{i}$ will be larger than $\ua_{i^{*}}$ and target $i^{*}$ will not be the attacker's best response. Hence, it is not possible that $p^{*}_{i^{*}} > p_{i^{*}}$, and thereby $(\b p, \b v)$ is an optimal valid defender strategy profile.

        Then, we will prove the time complexity of \cref{alg:exact} is $O(n^{4} \log n)$. In \cref{alg:exact}, there are totally $O(n(1 + \log \rv)) = O(n \cdot \log n)$ calls to \cref{alg:binary_judge}, and we have proved in \cref{theorem:binary_bound} that \cref{alg:binary_judge} works in $O(n)$.
        
        Besides, \cref{alg:exact} calls \cref{alg:exact_patroller} $O(n)$ times. In \cref{alg:exact_patroller}, on Lines 7 to 9, it takes $O(n \log n)$ time if we sort and maintain $\ua_{i}$ with heap. To optimize \cref{alg:swap_line}, we can first calculate the $n^{2}$ number of utility drop before swapping of tuple $(i_{\mr{outp}}, i_{\mr{outv}})$ before Line 11 in \cref{alg:exact_patroller} and maintain the utility drops of every $i_{\mr{outv}}$ with heap, which takes $O(n \log n)$ time. Every time we call \cref{alg:swap_line}, it takes $O(n)$ time to find the minimum utility drop by checking the minimum value of every $i_{\mr{outv}}$, and it takes $O(n \log n)$ to recalculate utility drop before swapping related to $i_{\mr{outv}}$ and $i_{\mr{outp}}$. Since we can move one villager from target $i_{\mr{outv}}$ to target $i_{\mr{outp}}$ only when $w_{i_{\mr{outp}}} > w_{i_{\mr{outv}}}$, there can be at most $n^{2}$ swaps. Combined with the fact there is at most $n$ times that $u_{\mr{cur}} - u_{\delta} < \pa_{\mr{max}}$ and $n$ times that $u_{\mr{cur}} - u_{\delta} < u_{\mr{next}}$, there are at most $O(n^{2} + 2n) = O(n^{2})$ rounds of iterations on Lines 11 to 44, and it takes $O(n^{3} \log n)$. Therefore, \cref{alg:exact_patroller} takes $O(n^{3} \log n)$ time. 
        
        Therefore, \cref{alg:exact} works in $O(n \log n \cdot n + n \cdot n^{3} \log n) = O(n^{4} \log n)$ time.
\end{proofof}

\section{Missing Algorithm in \cref{sec:variants}}
\label{app:missing_alg_variants}

In this section, we will present an algorithm to resolve RACPP with target-specific effectiveness, which is omitted in \cref{sec:variants}. Similar to \cref{alg:binary_judge}, the general idea is to greedily allocate the resources to other targets by maximizing the coverage each coverage can cover to ensure the best response. The formal procedure is shown in \cref{alg:binary_judge_refined}.

\begin{algorithm}[htbp]
    \caption{Checking whether a consistent strategy exists}
    \label{alg:binary_judge_refined}

    \textbf{Input}: Input instance $\I$, target $i^{*}$, strategy $(p_{i^{*}}$, $v_{i^{*}})$ on $i^{*}$

    \textbf{Output}: Whether a consistent strategy $(\b p, \b v)$ exists

    \begin{algorithmic}[1] 
        \STATE Compute $\ua_{i^{*}}$ using \eqref{equation:pv_to_ua} with $(p_{i^{*}},v_{i^{*}})$.
        \IF{$\exists i\in T, \ua_{i^{*}} < \pa_{i}$}
            \RETURN $\mathrm{False}$.
        \ENDIF
        \STATE Let $(\rp_{\mathrm{remain}}, \rv_{\mathrm{remain}}) \gets (\rp - p_{i^{*}}, \rv - v_{i^{*}})$.
        \FOR{$i \in \{0,1,\dots,n-1\} \setminus \{i^{*}\}$}
            \STATE Let $c_{\mathrm{min}, i} \gets \text{the minimum $c_i$ to ensure $\ua_{i} \leq \ua_{i^{*}}$}$.
            \STATE Let $c_{\mathrm{remain}, i} \gets c_{\mathrm{min}, i}$.
            \STATE Let $\delta_i \gets \min(c_{\mathrm{remain}, i}, \ev_{i})$.
        \ENDFOR
        \FOR{$t \in \{0, 1, \dots, \min(n, \rv_{\mathrm{remain}})-1\}$}
            \STATE Choose $i$ with $\text{the largest $\delta_i\ (i\in T)$}$.
            \STATE Let $c_{\mathrm{remain}, i} \gets c_{\mathrm{remain}, i} - \delta_{i}$.
            \STATE Let $\delta_{i} \gets \min(c_{\mathrm{remain}, i}, \ev_{i})$.
        \ENDFOR
        \RETURN $\sum_{i \neq i^{*}} c_{\mathrm{remain}, i} \leq \rp_{\mathrm{remain}} \cdot \ep$.
    \end{algorithmic}
\end{algorithm}

The algorithm maintains a variable $c_{\mathrm{remain}, i}$ (Line 7) representing the coverage still needed on target $i$ after allocating some villagers to it, and a variable $\delta_{i}$ (Line 8) representing the amount of coverage the next villager allocated to target $i$ can handle. For each villager, the algorithm allocates them to the target with the largest $\delta_{i}$, trying to maximize the coverage each villager can cover (Lines 9 to 12). Finally, the algorithm checks whether there are enough ranger efforts to cover the remaining coverage needs on all targets (Line 13).

\section{Missing Proofs in \cref{sec:variants}}
\label{app:missing_proofs_variants}

\subsection{Proof of \cref{lemma:binary_judge_correctness_refined}}
\label{appsub:proof_of_lemma_binary_judge_refined}

\BinaryJudgeCorrectnessRefined*

\begin{proofof}{\cref{lemma:binary_judge_correctness_refined}}
A valid defender strategy profile $(\b p, \b v)$ such that $p_{i^{*}} = p, v_{i^{*}} = v$ and $i^{*}$ is the attacker's best response should satisfy (a) $p_{i}\in \mathbb R_{\geq 0}$, (b) $v_{i} \in \mathbb{N}$, (c) $\sum_{i \in T}p_i\leq \rp$, (d) $\sum_{i \in T}v_i\leq \rv$, (e) $p_{i^{*}} = p$, (f) $v_{i^{*}} = v$, and (g) $\ua_{i^{*}} \geq \ua_{i}$ $(\forall i \in T)$.

    \textbf{Sufficiency.} After \cref{alg:binary_judge_refined}, we construct $(\b p, \b v)$ as 
    
    \begin{align*}
        p_{i} & = \left\{\begin{array}{ll}
            c_{\mathrm{remain}, i} / \ep & (i \neq i^{*}) \\
            p & (i = i^{*}),\end{array}\right. \\
        v_{i} & = \left\{\begin{array}{ll}
            \text{\# times $i$ is chosen on Line 10} & (i \neq i^{*}) \\
            v & (i = i^{*}).\end{array}\right.
    \end{align*}
    
    We first prove that if \cref{alg:binary_judge} returns True, the constructed $(\b p, \b v)$ is a valid defender strategy profile that satisfies (a) to (g). 

    After executing Lines 5 to 8, $c_{\mathrm{remain}, i} \geq 0$ since $c_{\mathrm{remain}, i}$ is set to $c_{\mathrm{min}, i}$. The property still holds after executing Lines 9 to 12 because $c_{\mathrm{remain}, i}$ is set to $c_{\mathrm{remain}, i} - \min(c_{\mathrm{remain}, i}, \ev_{i}) \geq 0$. Hence, $p_{i} =c_{\mathrm{remain}, i} / \ep \geq 0$ $(\forall i \in T \setminus \{i^{*}\})$. Combined with $p_{i^{*}} = p \geq 0$, (a) holds.

    Since $v_{i}$ $(\forall i \in T \setminus \{i^{*}\})$ is similar to a counter,  $v_{i} \in \mathbb{N}$ $(\forall i \in T \setminus \{i^{*}\})$. Combined with $v_{i^{*}} = v \in \mathbb{N}$, (b) holds.

    From Line 13, we know that $\sum_{i \neq i^{*}} c_{\mathrm{remain}, i} \leq \rp_{\mathrm{remain}} \cdot \ep = (\rp - p_{i^{*}}) \cdot \ep$. Since $p_{i} = c_{\mathrm{remain}, i} / \ep$ $(\forall i \in T \setminus \{i^{*}\})$, we can derive that $\sum_{i \in T}p_i \leq \rp$ and (c) holds.

    On Lines 9 to 12, there are at most $\rv_{\mathrm{remain}}$ iterations, and only one of the targets with the largest $\delta_{j}$ can be chosen within one iteration. Hence, $\sum_{i \neq i^{*}} v_{i} \leq \rv_{remain} = \rv - v_{i^{*}}$. Therefore, $\sum_{i \in T}v_i \leq \rv$ and (d) holds.

    (e) and (f) hold since we set $p_{i^{*}}$ and $v_{i^{*}}$ to $p$ and $v$.

    On Line 6, we calculate a minimum coverage $c_{i}$ as $c_{\mathrm{min}, i}$ to ensure $\ua_{i} \leq \ua_{i^{*}}$. After executing every iteration on Lines 9 to 12, $c_{\mathrm{remain}, i} + v_{i} \cdot \ev_{i} \geq c_{\mathrm{min}, i}$. Hence, $p_{i} \cdot \ep + v_{i} \cdot \ev_{i} \geq c_{\mathrm{min}, i}$ $(\forall i \in T \setminus \{i^{*}\})$ and (g) holds.
    
    Therefore, there exists a valid defender strategy profile $(\mathbf p, \mathbf v)$ such that $p_{i^{*}} = p, v_{i^{*}} = v$ and $i^{*}$ is the attacker's best response if \cref{alg:binary_judge_refined} returns True.

    \textbf{Necessity.} We will prove the necessity of \cref{alg:binary_judge_refined} by contradiction. Assume that there exists a valid defender strategy profile $(\b p,\b v)$ such that $p_{i^{*}} = p, v_{i^{*}} = v$ and $i^{*}$ is the attacker's best response, while \cref{alg:binary_judge_refined} returns False.
    
    Although \cref{alg:binary_judge_refined} returns False, it generates a defender strategy $\b v'$ for villagers, where $v'_{i}$ $(\forall i \in T \setminus \{i^{*}\})$ is the number of times $i$ is chosen on Line 10 and $v'_{i^{*}} = v$. The strategy $\b v'$ satisfies properties (b), (d) and (f) for the reasons similar to the sufficiency part.
    
    Without loss of generality, we can assume both $\b v$ and $\b v'$ allocate all villagers to the targets. Hence, we know that $\sum_{i \neq i^{*}} v_{i} = \sum_{i \neq i^{*}} v'_{i}$.
    
    On Lines 9 to 12, \cref{alg:binary_judge_refined} distributes villages to targets with maximum $\delta_{i}$, which is the effect one villager can play on target $i$. Hence, $\b v'$ guarantees a maximum $\sum_{i \neq i^{*}}(v'_{i} \cdot \ev_{i} - \cw_{i}(\b v'))$ and we can know that $\sum_{i \neq i^{*}}(v_{i} \cdot \ev_{i} - \cw_{i}(\b v)) \leq\sum_{i \neq i^{*}}(v'_{i} \cdot \ev_{i} - \cw_{i}(\b v'))$.
    
    Therefore, after villager distributions are set, the sum of coverage needed to by filled with rangers over all targets $\sum_{i \neq i^{*}}(c_{\mathrm{min}, i} - v'_{i} \cdot \ev_{i} + \cw_{i}(\b v')) \leq \sum_{i \neq i^{*}}(c_{\mathrm{min}, i} - v_{i} \cdot \ev_{i} + \cw_{i}(\b v))$. Since \cref{alg:binary_judge_refined} returns False, we derive that $\sum_{i \neq i^{*}}(c_{\mathrm{min}, i} - v'_{i} \cdot \ev_{i} + \cw_{i}(\b v')) > (\rp - p) \cdot \ep$. Therefore, we can conclude that $(\rp - p) \cdot \ep < \sum_{i \neq i^{*}}(c_{\mathrm{min}, i} - v_{i} \cdot \ev_{i} + \cw_{i}(\b v'))$, which means there are no enough ranger efforts to construct a $\b p$ as the strategy for ranger in the valid defender strategy profile $(\b p, \b v)$. We get a contradiction.
\end{proofof}

\begin{figure*}[t]
    \centering
    
    \begin{subfigure}{0.48\textwidth}
        \centering
        \includegraphics[width=\linewidth]{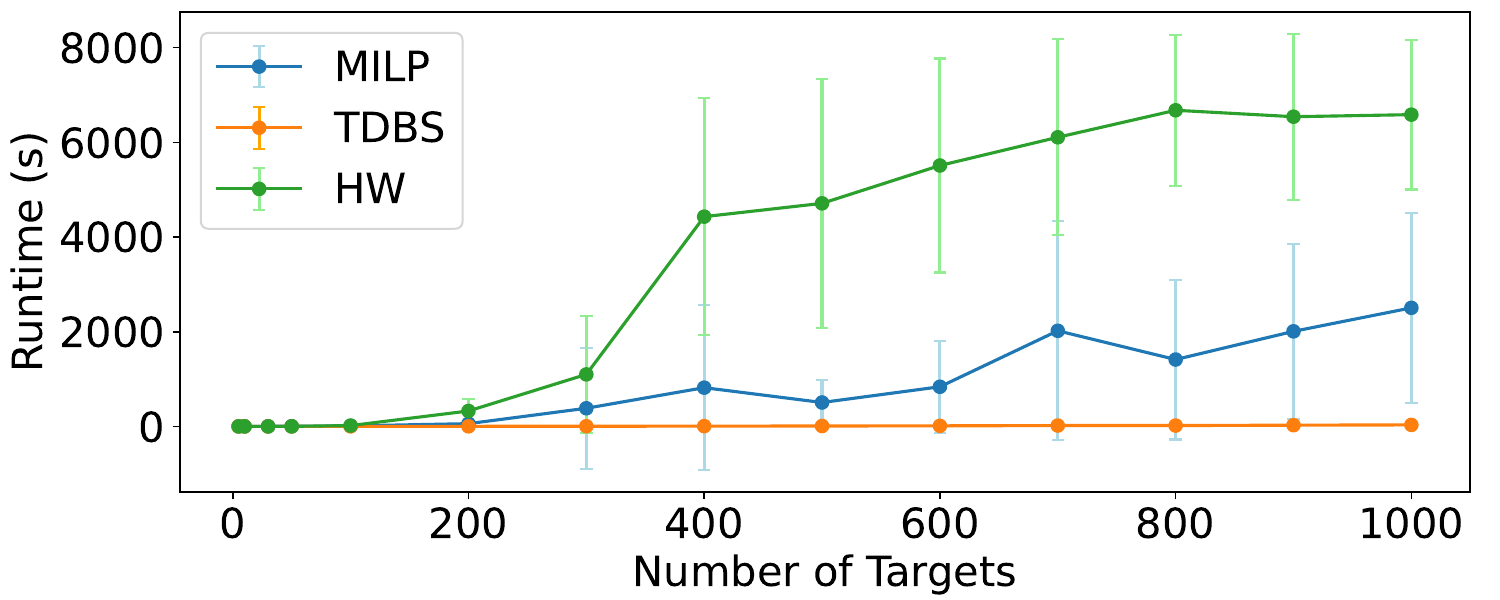}
        \caption{runtime with different $n$ and $\rv = 2 \cdot \rp = 2 \cdot \lfloor \frac{n}{3} \rfloor$}
        \label{subfigure:n-runtime2}
    \end{subfigure}
    \hfill
    \begin{subfigure}{0.48\textwidth}
        \centering
        \includegraphics[width=\linewidth]{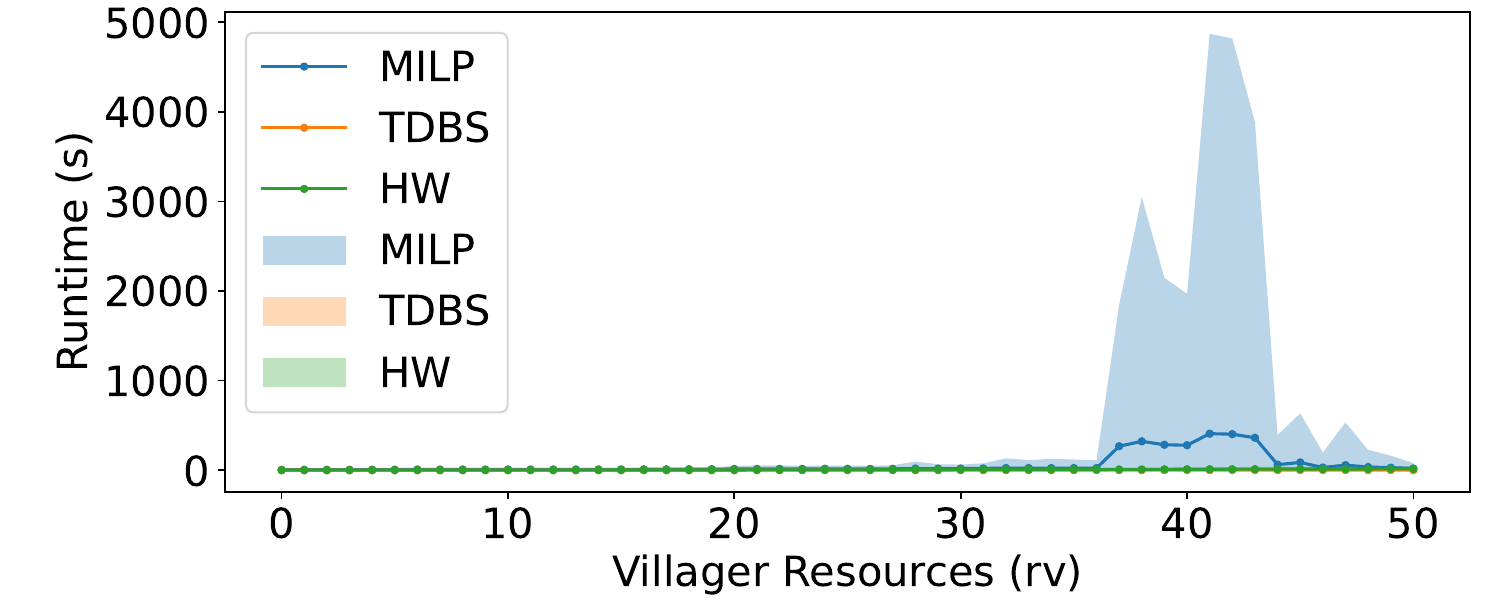}
        \caption{runtime with $n = 100$ and different $\rp,\rv$}
        \label{subfigure:rv-runtime2}
    \end{subfigure}
    
    \caption{Average runtime of MILP, TDBS, and HW over 30 runs under different combinations of $(n,\rp,\rv)$ where $\rv = 2 \cdot \rp$. The error bars indicate the standard deviation. The shaded areas represent the ranges from the minimum to the 97th percentile. We limit the maximum runtime for MILP in \cref{subfigure:rv-runtime2} to 7200 seconds. 
    }
    \label{fig:performance2}
\end{figure*}

\subsection{Proof of \cref{theorem:NP_hardness}}
\label{appsub:proof_of_theorem_NPHardness}

\NPHardness*

\begin{proofof}{\cref{theorem:NP_hardness}}
    Consider the case where $n = 2$, $\rp = 0$, $\ra_{i} = -\pd_{i} = a$, $\rd_{i} = -\pa_{i}= b$ for all $i \in T$, and $\sum_{j \in V} e_{j} < 2$.

To start with, we need to prove that in this case, maximum utility can be achieved when $e_{j}$ are uniformly distributed to the 2 targets if possible. We assume that there exists a partition which splits $V$ into two sets $S_{1}$ and $S_{2}$ where $\sum_{i \in S_{1}} e_{i} = \sum_{i \in S_{2}} e_{i}$, $S_{1} \cap S_{2} = \emptyset$ and $S_{1} \cup S_{2} = V$, while maximum utility is reached with $S'_{1}$ and $S'_{2}$ respectively allocated to the 2 targets, where $\sum_{i \in S'_{1}} e_{i} > \sum_{i \in S'_{2}} e_{i}$, $S'_{1} \cap S'_{2} = \emptyset$ and $S'_{1} \cup S'_{2} = V$. Combined with $\sum_{j \in V} e_{j} < 2$, we can get that $c'_{1} > c_{1} = c_{2} > c'_{2}$, thus $\udp_{1} > \ud_{1} = \ud_{2} > \udp_{2}$ and $\uap_{1} < \ua_{1} = \ua_{2} < \uap_{2}$. Therefore, the attacker will choose to attack target 2 with best response, and thereby defenders will obtain utility $U_{2}^{\prime d}$ with $S'_{1}$ and $S'_{2}$. Similarly, defenders will get utility $U_{1}^{d} = U_{2}^{d}$ with $S_{1}$ and $S_{2}$. Since $\ud_{1} = \ud_{2} > \udp_{2}$, there is a contradiction. Hence, if $e_{j}$ can be divided into two sets with equal sums, maximum utility will be achieved with the corresponding strategy profile.

We show that the optimization program is NP-Hard using a reduction from Partition Problem: Given a collection of positive integers $S = \{a_{1}, a_{2}, \ldots, a_{m}\}$, decide if there is a partition of $S$ to $A$ and $B$ such that $\sum_{a \in A} a = \sum_{b \in B} b$. Let $sum = \sum_{a_{i} \in S} a_{i}$, $|V| = m$ and $e_{j} = a_{j} / sum$. Then, solve the RACPP problem in this case. Respectively compute the sum of effectiveness allocated to the 2 targets within time complexity $O(m)$, and if the two sums are equal, there is a partition in the Partition Problem; otherwise, there are no such partition.

Hence, Partition Problem $\leq_{k}$ RACPP with villager-specific effectiveness, and computing the maximum utility and the optimal valid strategy profile is NP-Hard.
\end{proofof}

\section{Additional Experiments}
\label{app:additional_experiments}

In reality, the number of villagers and the number of rangers are not always equal. To further validate our findings regarding performance, we conducted a series of additional experiments with a specific configuration where $rv = 2 \cdot rp$, in addition to maintaining the same experiment setup as described in \cref{sec:experiments}. This configuration was chosen to simulate scenarios where the number of villagers exceeds the number of rangers, aligning with practical considerations in wildlife conservation efforts.

\paragraph{Performance with different $n$.} We let $n = \{5, 10, 30,$ $50, 100, 200, 300, 400, 500, 600, 700, 800, 900, 1000\}$, and let $\rv = 2 \cdot \rp = 2 \cdot \lfloor \frac{n}{3} \rfloor$. As shown in \cref{subfigure:n-runtime2}, the runtime of TDBS is significantly lower than the other two algorithms. 

\paragraph{Performance with different $\rp,\rv$.} We then fix $n=100$ and let $\rv = 2 \cdot \rp$ be $\{0,1,\dots,50\}$. As shown in \cref{subfigure:rv-runtime2}, runtimes of TDBS and HW are stable with different $\rp$ and $\rv$, while the performance of MILP is highly unstable.

The results confirm the conclusions in the main paper. Specifically, TDBS is significantly faster with sufficient accuracy, and TDBS and HW maintain greater stability.

\section{Additional Case Study}
\label{app:additional_case_study}

\begin{figure*}[t]
  \centering
  \includegraphics[width=0.285\linewidth]{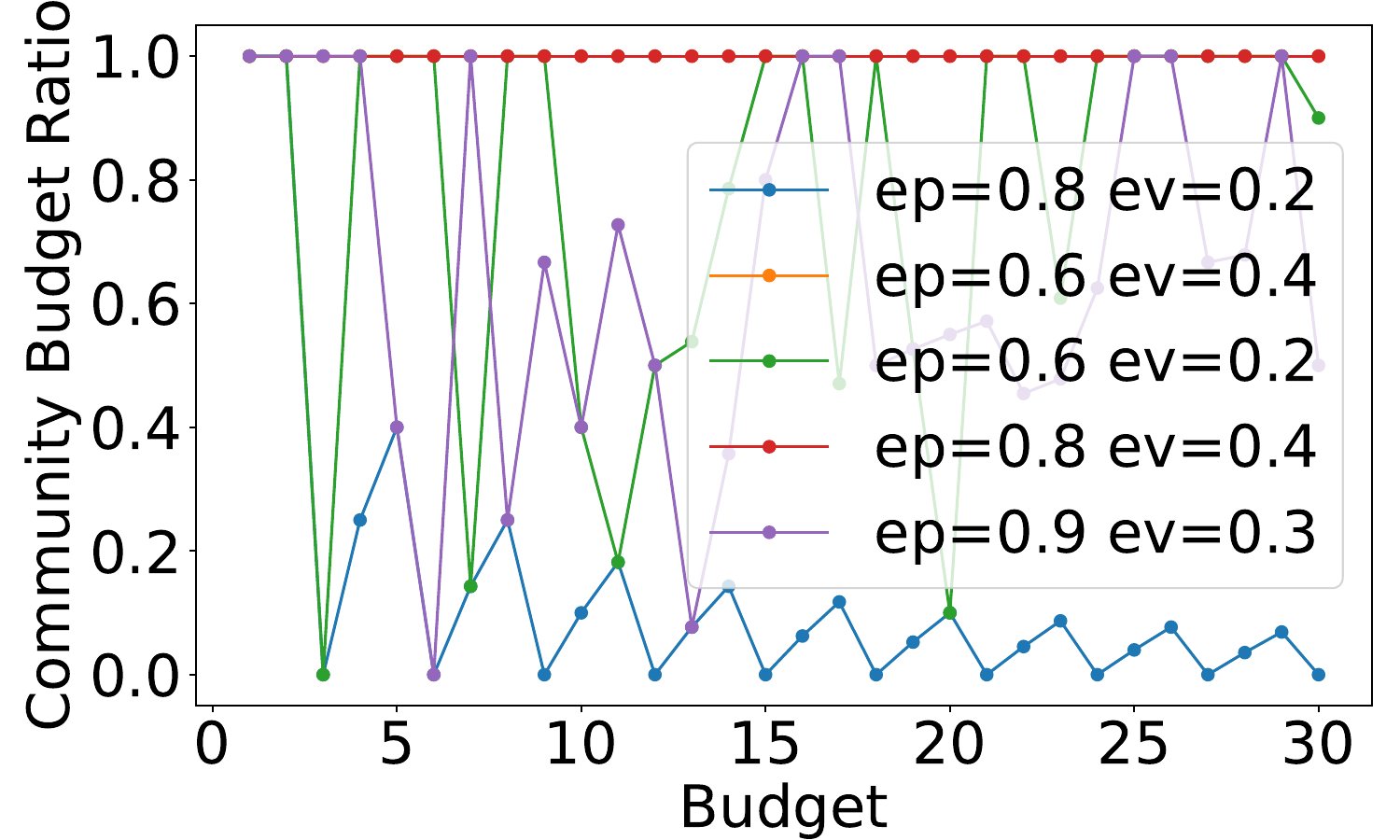}
  \caption{Budget allocation with more settings}
  \label{fig:budget_app}
\end{figure*}

\begin{figure*}[t]
    \centering
    
    \begin{subfigure}{0.285\textwidth}
      \centering
      \includegraphics[width=\linewidth]{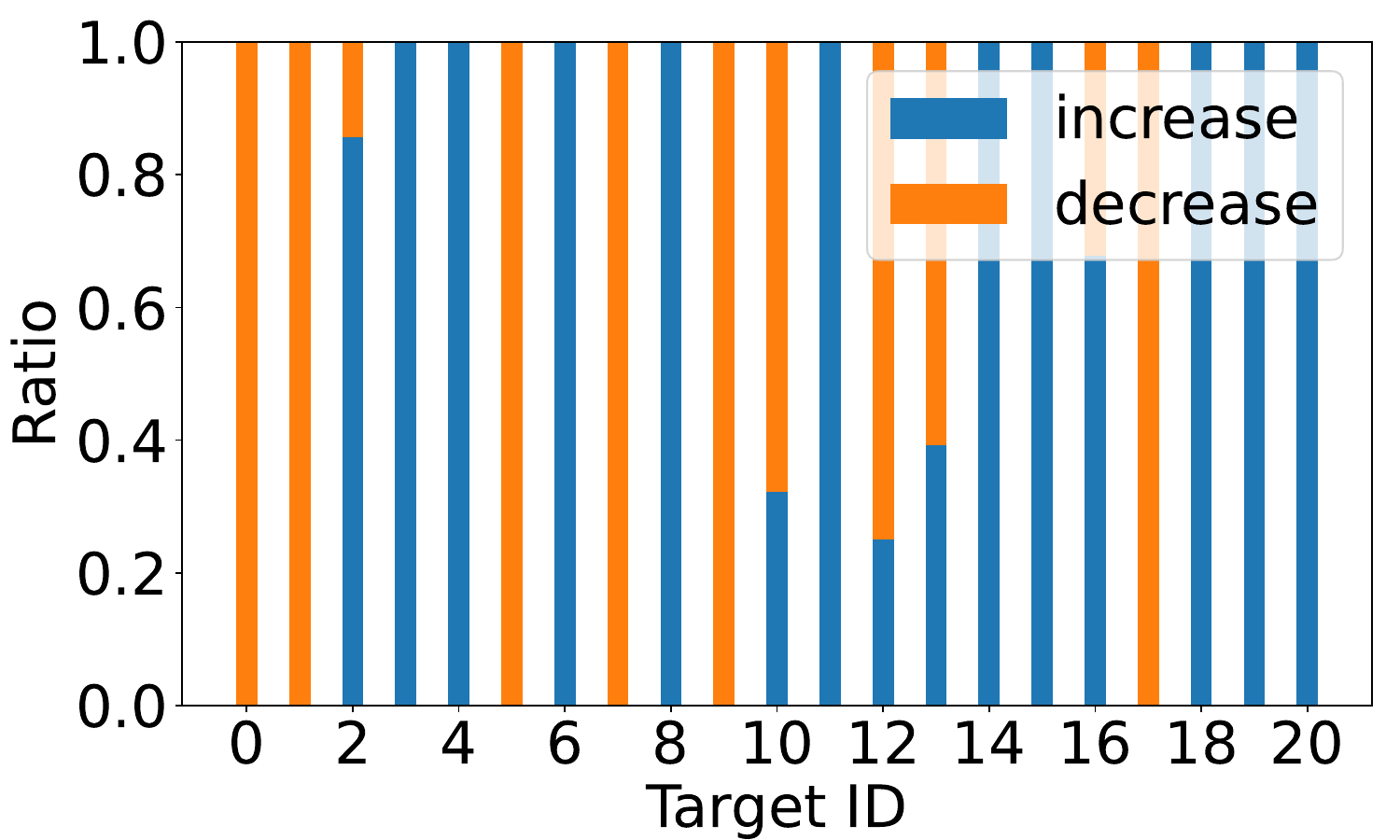}
      \caption{Ratio of increase and decrease}
      \label{subfigure:ratio_terrain}
    \end{subfigure}
    \hfill
    \begin{subfigure}{0.285\textwidth}
      \centering
      \includegraphics[width=\linewidth]{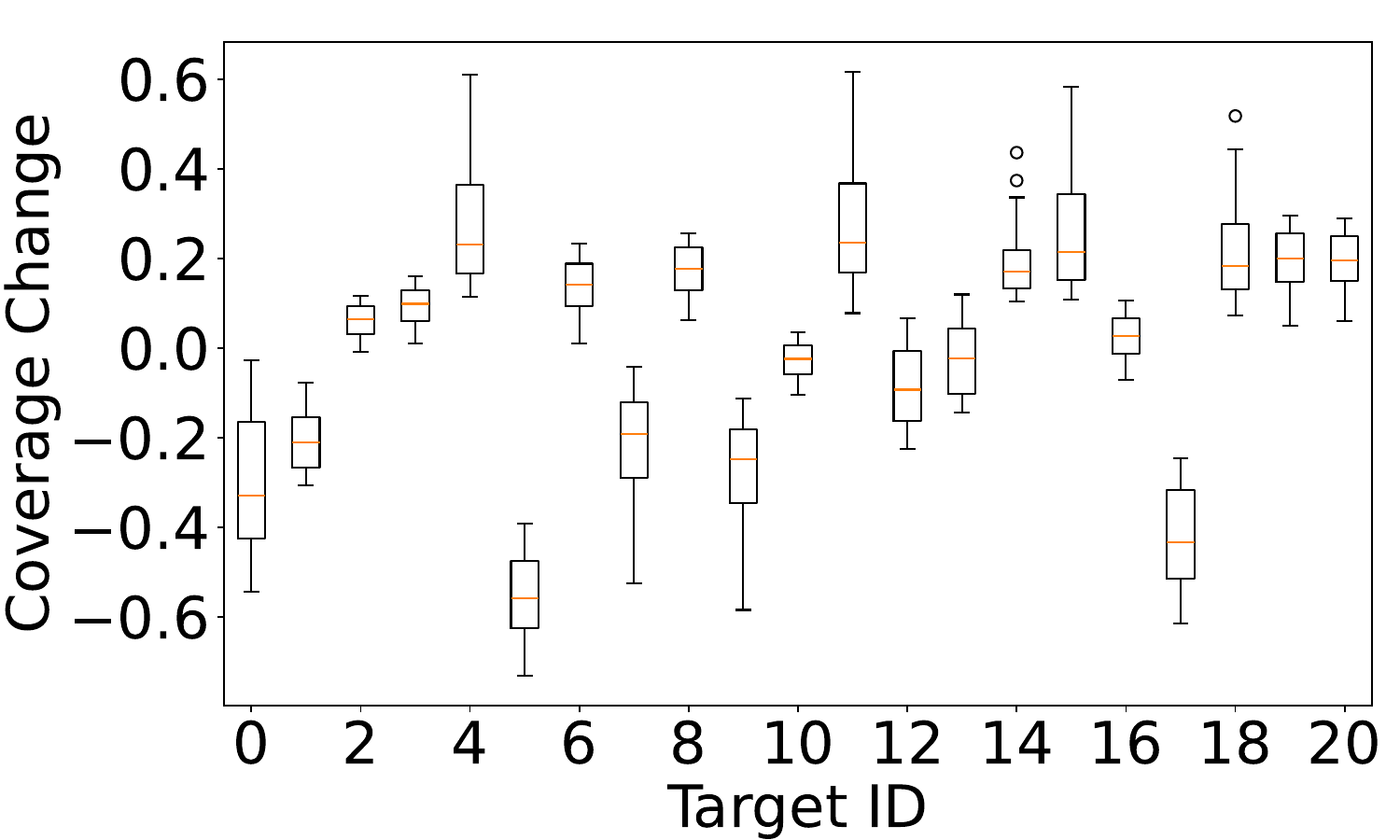}
      \caption{Coverage change}
      \label{subfigure:coverage_change_terrain}
    \end{subfigure}
    \hfill
    \begin{subfigure}{0.285\textwidth}
        \centering
        \includegraphics[width=\linewidth]{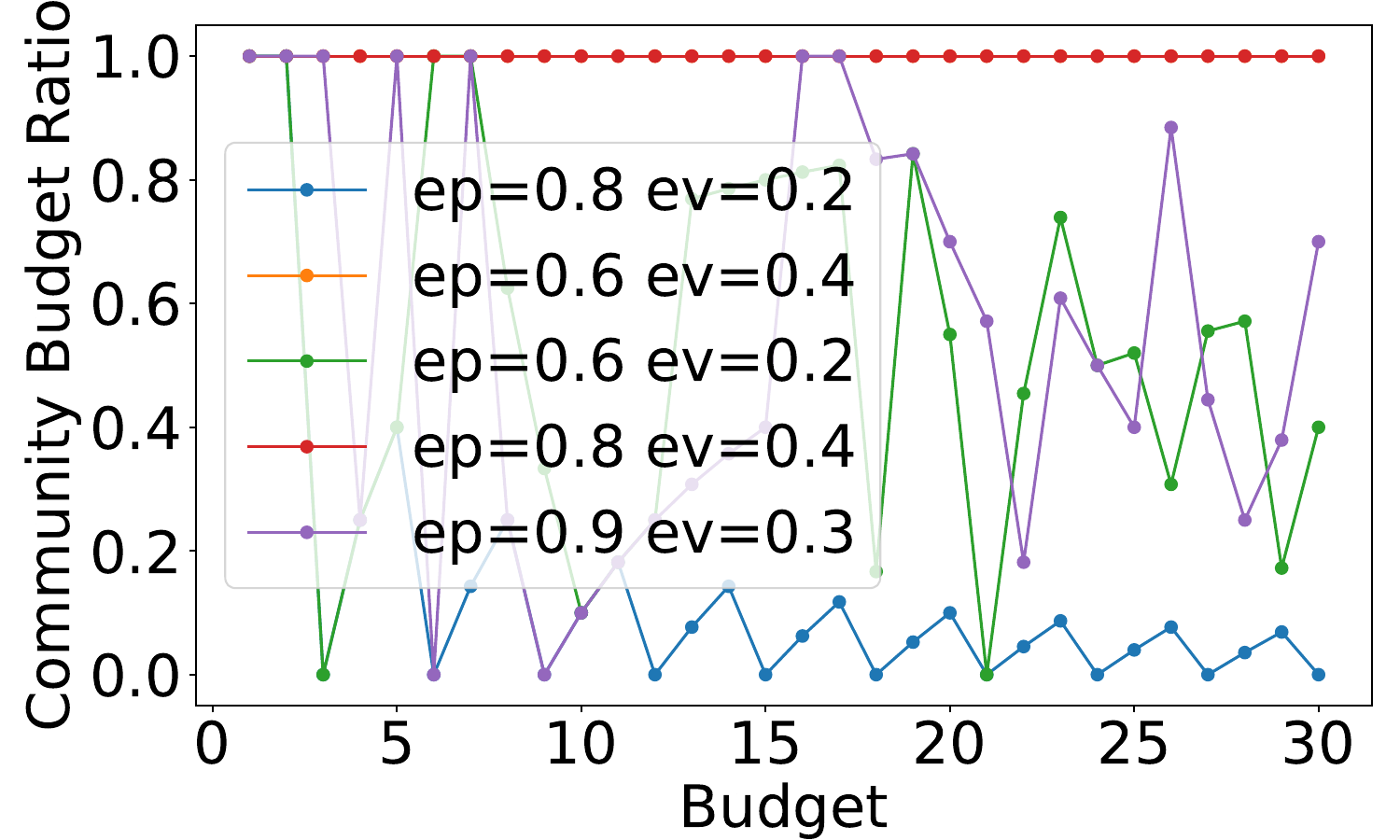}
        \caption{Budget allocation}
        \label{subfigure:budget_terrain}
    \end{subfigure}

    \caption{Case study results considering terrain.}
    \label{fig:case_study_terrain}
\end{figure*}

Distributions of three wildlife species are provided by domain experts. The distributions are processed, weighted and scaled to get the value $\ra_{i}$ of every target, which is shown is \cref{table:target_value}.

\begin{table}[htbp]
    \centering
    \begin{tabular}{|c|c|c|c|c|c|}
        \hline
        Target & Value & Target & Value & Target & Value \\
        \hline
        0 & 6.83 & 7 & 5.64 & 14 & 6.90 \\
        \hline
        1 & 6.18 & 8 & 5.92 & 15 & 6.38 \\
        \hline
        2 & 5.18 & 9 & 5.11 & 16 & 4.52 \\
        \hline
        3 & 5.84 & 10 & 4.35 & 17 & 5.59 \\
        \hline
        4 & 6.23 & 11 & 6.18 & 18 & 5.59 \\
        \hline
        5 & 4.18 & 12 & 6.93 & 19 & 5.46 \\
        \hline
        6 & 4.98 & 13 & 6.76 & 20 & 5.70 \\
        \hline
    \end{tabular}
    \caption{Reward for poacher $\ra_{i}$ used in the case study}
    \label{table:target_value}
\end{table}

For extra budget allocation, we calculate the optimal plan of allocating the extra money with more $(\ep, \ev)$ settings. The results are demonstrated in \cref{fig:budget_app}. We can observe that with $\ep / \ev$ equal to 3, the preference to rangers or villagers fluctuates when budget increases.

In practice, terrain has a great impact on defense effectiveness. There is a basic domain knowledge that finding snares on targets with higher slope variance is easier for defenders. Hence, we repeat the case study in \cref{sec:case_study} with slope taken into consideration.

First, we collect the terrain statistics of the forest farm, compute the slope variance on every target, and classify the targets into three categories (high, average and low slope variance). Then, every $(\ep, \ev)$ tuple is modified according to the classification. For every $(\ep, \ev)$ tuple, we set 
\begin{equation*}
        e^{\mr{d}}_{i} =\left\{\begin{array}{ll}
            e^{\mr{d}} + 0.1 & (i \text{ with high slope variance}) \\
            e^{\mr{d}}  & (i \text{ with average slope variance})\\
            e^{\mr{d}} - 0.1 & (i \text{ with low slope variance}),
        \end{array}\right.
    \end{equation*}
for $\mr{d} \in \{\mr{p}, \mr{v}\}$. 

Then, we generate strategy suggestions and budget allocation with the terrain considered $(\ep, \ev)$. The results are shown in \cref{fig:case_study_terrain}, which is more applicable because key factors in reality is taken into account. We can see from \cref{subfigure:budget_terrain} that the slight differences of defense effectiveness among targets do not affect budget allocation when $\ep/\ev$ is not close to the cost ratio.

\end{document}